\documentclass[fleqn,usenatbib]{mnras}

\usepackage{newtxtext,newtxmath}

\usepackage[T1]{fontenc}
\usepackage{ae,aecompl}
\usepackage{xspace}

\usepackage{graphicx}	
\usepackage{amsmath}	

\newcommand{\bs}[1]{\boldsymbol{#1}}

\usepackage{bm}
\let\vec\bm%

\usepackage{mleftright}
\let\left\mleft%
\let\right\mright%

\newcommand{\be}{\begin{equation}}
\newcommand{\en}{\end{equation}}

\newcommand{\pder}[2]{\frac{\partial{#1}}{\partial{#2}}}
\newcommand{\pdder}[2]{\frac{\partial^2{#1}}{\partial{#2}^2}}

\newcommand{\der}[2]{\frac{\mathrm{d}#1}{\mathrm{d}#2}}

\newcommand{\Dx}{\Delta{}x}

\newcommand{\f}{\frac}

\newcommand{\ex}{\vec{e}_x}
\newcommand{\ey}{\vec{e}_y}
\newcommand{\ez}{\vec{e}_z}

\newcommand{\para}{\parallel}

\renewcommand{\b}{\bb{b}}
\newcommand{\rhos}{\rho_{\mathrm{s}}}
\newcommand{\Ts}{T_{\mathrm{s}}}

\newcommand{\dvx}{\delta \varv_{x}}
\newcommand{\dvz}{\delta \varv_{z}}

\newcommand{\dt}{\frac{\delta T}{T}}
\newcommand{\drho}{\frac{\delta \rho}{\rho}}
\newcommand{\dA}{\f{\delta A}{B}}

\newcommand\bb[1]{\mbox{\boldmath{$#1$}}}

\newcommand{\va}{\varv_\mathrm{a}}

\newcommand{\cs}{c}

\title[Impact of magnetic fields on cold streams]
{The impact of magnetic fields on cold streams feeding galaxies}

\author[Thomas Berlok and Christoph Pfrommer]{
Thomas Berlok\thanks{E-mail: tberlok@aip.de} and
Christoph Pfrommer
\\
Leibniz-Institut f{\"u}r Astrophysik Potsdam (AIP), An der Sternwarte 16, D-14482 Potsdam, Germany
}

\date{Accepted XXX. Received YYY; in original form ZZZ}

\pubyear{2019}

\begin{document}
\label{firstpage}
\pagerange{\pageref{firstpage}--\pageref{lastpage}}
\maketitle


\begin{abstract}
High redshift, massive halos are observed to have sustained high star
formation rates, which require that the amount of cold gas in the halo is
continuously replenished. The cooling time scale for the hot virialized halo
gas is too long to provide the source of cold gas.  Supersonic, cold streams
have been invoked as a mechanism for feeding massive halos at high redshift
and delivering the cold gas required for continued star formation at the rates
observed. This mechanism for replenishing the cold gas reservoir is motivated
by some cosmological simulations.  However, the cold streams are likely to be
subject to the supersonic version of the Kelvin-Helmholtz instability (KHI),
which eventually leads to stream disruption.  Cosmological simulations have
yet to obtain the spatial resolution required for understanding the detailed
stability properties of cold streams.  In this paper, we consider instead an
idealized model of magnetized cold streams that we spatially resolve.  Using
linear theory we show how magnetic fields with dynamically important field
strengths do not inhibit the KHI but rather \emph{enhance} its growth rate. We
perform nonlinear simulations of magnetized stream disruption and find that
magnetic fields can nevertheless increase stream survival times by suppressing
the mixing rate of cold gas with the circumgalactic medium.  We find that
magnetic fields can allow streams to survive $\sim 2-8$ times longer and,
consequently, that streams $\sim 2-8$ times thinner can reach the central
galaxy if the magnetic field strength is $\sim 0.3-0.8 \mu$G.
\end{abstract}

\begin{keywords}
galaxies: high-redshift -- instabilities -- (magnetohydrodynamics) MHD
\end{keywords}

\section{Introduction}
\label{sec:intro}
High redshift ($z\sim2-3$), massive galaxies  residing in $\sim10^{12}M_\odot$
dark matter halos are observed to have high star formation rates of $\sim 100
M_\odot  \mathrm{yr}^{-1}$ \citep{Genzel2006,Schreiber2006}.  This peak epoque
of cosmic star formation requires a continuous  source of new cold gas. The
galactic cold gas reservoir, which is needed to  fuel intense star formation,
would otherwise be depleted on a time scale shorter than observed. Determining
whether the star formation is sustained by recycling of gas, in situ cooling
or whether it is provided by an external source is an important topic in
galaxy formation (see e.g. \citealt{Somerville2015}, \citealt{Naab2017} and
\citealt{Dayal2018} for reviews of galaxy formation).

Theories of structure formation predict that the halo gas is heated to the
virial temperature, $T_\mathrm{v}$, by an  accretion shock when the halo mass
is  $\gtrsim 10^{12} M_\odot$  \citep{Birnboim2003}. Such hot ($T_\mathrm{v}
\gtrsim 10^6$ K) virialized gas has a prohibitively long cooling time  for
making in situ cooling a plausible source of cold gas
\citep{Rees1977,White1978,Silk1977}. An alternative source of cold gas could
be mergers but observations have indicated that star formation takes place in
a large rotating disk, disfavoring this scenario as well
\citep{Genzel2006,Stark2008}.

In the cold stream model of galaxy formation in massive halos
\citep{Dekel2006,Dekel2009}, the central galaxy in a massive hot halo is
instead fed with cold gas that streams along filaments in the cosmic web and
penetrates down into the central galaxy.  Such cold streams have been observed
in some cosmological simulations  (e.g.
\citealt{Keres2005,Ocvirk2008,VanDeVoort2012a,Goerdt2015}), where they  are
able to provide the halo with  $\sim 100 M_\odot \mathrm{yr}^{-1}$ cold gas,
in rough agreement with the  observed star formation rate.

State-of-the-art zoom-in cosmological simulations are however currently in
disagreement on whether the cold stream model outlined above works. While some
simulations do find cold streams that feed the central galaxy, others find
that the streams are disrupted and heated inside $\lesssim 0.25-0.5
R_\mathrm{v}$ (where $R_\mathrm{v}$ is the virial radius,
\citealt{Nelson2013}). The discrepancy between different cosmological
simulations has been attributed to numerical differences in simulations
employing smoothed particle hydrodynamics (SPH), Eulerian adaptive mesh
refinement (AMR), and a quasi-Lagrangian finite volume approach, i.e., a
moving mesh \citep{Nelson2013}.

A key question that needs to be addressed for the cold stream model is whether
streams can in fact penetrate sufficiently far into the halo or whether they
will be disrupted by hydrodynamical instabilities during their propagation.
The question cannot be directly answered with current-day zoom-in cosmological
simulations, because such simulations have not attained the required spatial
resolution to accurately capture potentially disrupting hydrodynamical
instabilities.  Cold streams have been observed to have widths of 1-10 per
cent of the virial radius while the true width could be even smaller since
those values are close to the numerical resolution limit. For a halo with
$\sim 50$ kpc virial radius, this means that the thinnest streams are expected
to have width of $\sim 1$ kpc \citep{Mandelker2019} or smaller. They are
therefore not well resolved in cosmological simulations, despite recent
progress where the spatial resolution in the circumgalactic medium (CGM) is
enhanced to $\lesssim 1$ kpc
\citep{Suresh2019,Hummels2018,VandeVoort2019,Peeples2018}.  We note that the
resolution requirement for studying the multi-phase gas of the CGM might be a
general hindrance for obtaining high fidelity cosmological simulations of the
CGM in the near future, as some studies show that it is necessary to resolve
the gas all the way down to sub-pc scales \citep{McCourt2018,Sparre2019}.

As an alternative to studying cold flows in full cosmological simulations,
\citet{Mandelker2016,Padnos2018,Mandelker2019} have pursued the problem of
stream disruption using analytical theory and idealized simulations. They
found that the main disruption mechanism for a cold, dense stream propagating
through a less dense but hotter background, is the Kelvin-Helmholtz
instability (KHI). For supersonic propagation, the KHI can manifest itself as
sound waves reflecting and growing in amplitude inside of the stream (called
reflective or body modes, \citealt{Payne1985,Hardee1988}). This type of KHI
differs from the familiar textbook version  (e.g.
\citealt{Chandrasekhar1961,drazin2004hydrodynamic}) and the resulting breakup
of the stream does not contain the characteristic cat's eye vortex  (see Figs.
5 and 15 in \citealt{Padnos2018} for a comparison between the two types of
instability).

\citet{Mandelker2019} found that mixing of the stream with the background CGM
(and deceleration of the stream) occurs much faster in three dimensions in
comparison to two dimensions. It was concluded that the KHI can lead to
disintegration of cold streams before they reach the galaxy residing at the
center of a massive halo. For typical parameters, estimates showed that
streams with width less than $\sim 1-10$ per cent of the virial radius will be
disintegrated \citep{Mandelker2019}.  The deceleration and consequent
conversion of kinetic to thermal energy by the KHI could be a viable power for
observed Ly$\alpha$ blobs \citep{Dijkstra2009,Goerdt2012,Wisotzki2018}.

The analyses in \citet{Mandelker2016}, \citet{Padnos2018} and
\citet{Mandelker2019}, while extremely comprehensive, considered the KHI using
ideal hydrodynamics, and thus neglected potentially important influences on
stream behavior from physical effects such as radiative cooling, thermal
conduction, gravity and magnetic fields. In this paper, we present an
idealized model of magnetized cold streams in order to understand the impact
of magnetic fields on cold streams feeding galaxies. This inclusion of
magnetic fields in the analysis is motivated by widespread observations of
dynamically important magnetic fields in a large variety of astrophysical
systems \citep[see][for reviews on generation, evolution and observation of
magnetic fields]{Durrer2013,Han2017, Subramanian2019}.

Numerical simulations of galaxy formation have shown that a small-scale dynamo
can amplify magnetic seed fields in idealised setups to observed field
strengths \citep{Wang2009,Pakmor2013,Rieder2016,Rieder2017a,Steinwandel2019}
potentially aided by a cosmic-ray driven dynamo
\citep{Hanasz2004,Pakmor2016b}. Cosmological simulations show that tiny
magnetic seed fields grow exponentially by a small-scale dynamo driven by the
gravitational collapse and by supernova feedback until it saturates at around
redshift $z\simeq4$ with a magnetic energy of about 10 per cent of the kinetic
energy in the interstellar medium of the protogalaxy with typical strengths of
10 to 50~$\umu$G
\citep{Beck2012,Pakmor2014,Marinacci2015,Pakmor2017,Rieder2017b,Pakmor2018}.

Amplification takes place in two phases: in the kinematic phase the magnetic
field grows exponentially, primarily at the smallest nonresistive scale. This
is followed by the nonlinear phase, which transfers the magnetic energy toward
larger scales until the dynamo saturates on the turbulent forcing scale
\citep{Schober2013}. Exponential growth occurs irrespective of the seeding
mechanism, which can range from primordial magnetogenesis to battery processes
during the proto-galaxy formation \citep[modelled by injecting dipole-shaped
magnetic fields in][]{Beck2013}. After redshift $z\simeq6$, strong galactic
outflows driven by supernova and active galactic nucleus feedback transport
metals and magnetic fields into the CGM and cosmological filaments so that the
magnetization reaches levels of 0.1--1~$\umu$G close to galactic halos in
simulations \citep{Vazza2015,Marinacci2018, Nelson2019}.

Observationally, there is little known about magnetic field strengths in
cosmic filaments, cold streams and the CGM of massive halos, in particular at
the peak of cosmic star formation around $z\simeq2$. Using strong
gravitational lensing of polarized background quasars by galaxies enables
probing magnetic fields in cosmologically distant galaxies. Using differential
polarization properties (Faraday rotation and fractional polarization) of such
a lensing system at $z=0.44$, \citet{Mao2017} detected coherent $\umu$G
magnetic fields in the lensing disk galaxy.  The observation of radio halos in
nearby edge-on galaxies provides direct evidence that galactic outflows
transport magnetic fields several kpcs into the halo
\citep{Tullmann2000,Heesen2018,Stein2019}, and motivates our study of how
magnetic fields impact cold streams feeding galaxies.

The paper is divided as follows: we present idealized models of cold streams
in two- and three dimensions in Section~\ref{sec:ideal-model}.  We analyze
these idealized models using linear theory in Section~\ref{sec:lin_theory},
and find that fields modify the KHI in the supersonic regime. Remarkably, we
find that magnetic fields can \emph{increase} the growth rate of the
supersonic KHI with respect to the hydrodynamic case. The non-linear dynamics
of cold streams is then studied in Section~\ref{sec:simulations} by performing
numerical simulations. We present two-dimensional (2D) simulations of cold
streams in Section~\ref{sec:2D} and proceed to study the more realistic,
three-dimensional (3D) models in Section~\ref{sec:3D}. Here we find that
magnetic fields can \emph{suppress} mixing of cold streams with the CGM during
the nonlinear evolution of the instability.  We assess the differences found
between 2D and 3D in Section~\ref{sec:2D-vs-3D} and discuss the astrophysical
implications of our study in Section~\ref{sec:astro-impli}, in particular how
magnetic fields could allow streams that are $\sim 2-8$ thinner to reach the
central galaxy.  We conclude in Section~\ref{sec:conclusion} by summarizing
and pointing to future work.

\section{Idealized, Magnetized, Cold Stream Model}
\label{sec:ideal-model}

\begin{figure}
\centering
\includegraphics[width=1\columnwidth]{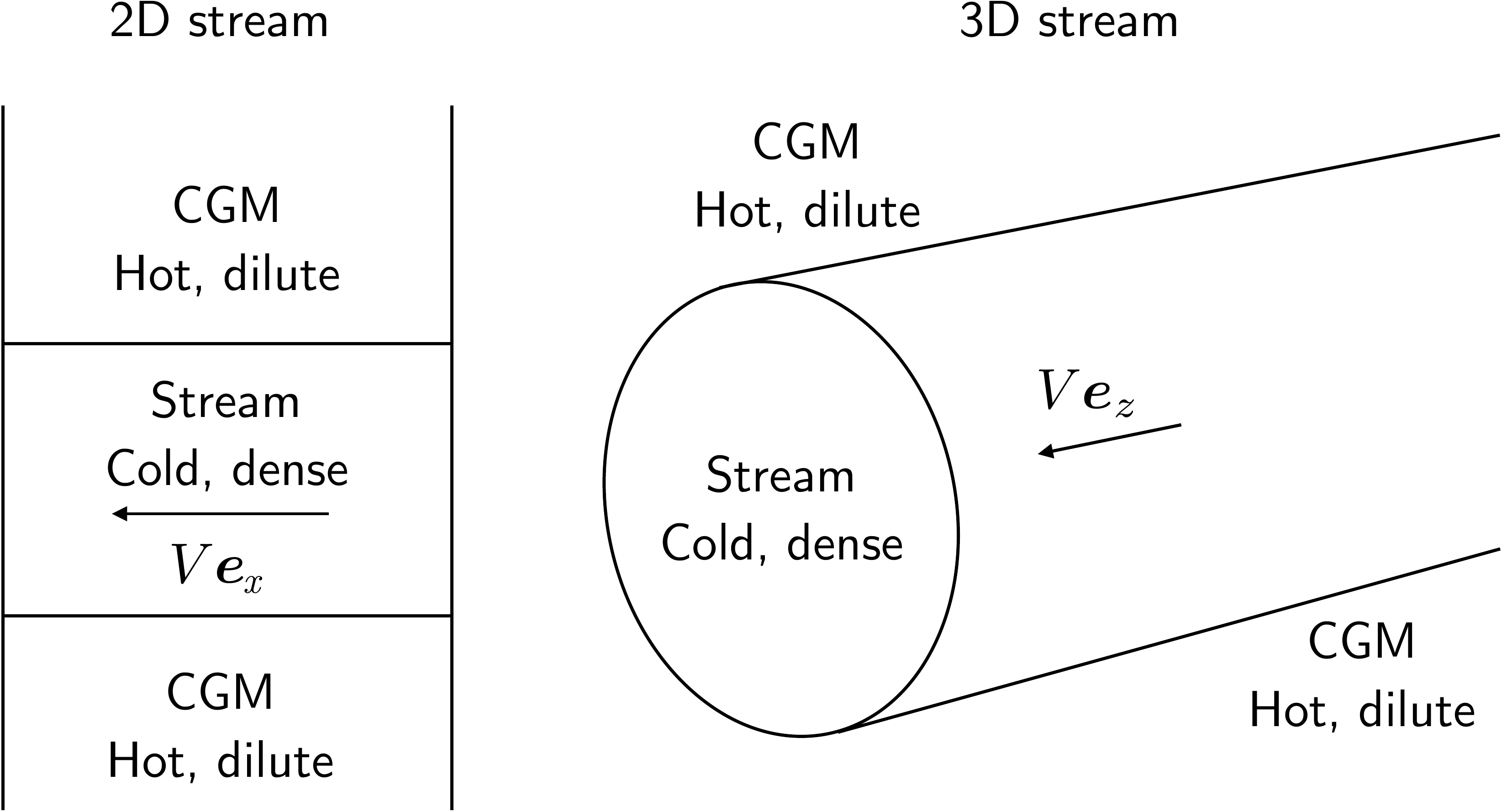}
\caption{The two models of cold, dense streams of gas moving with
supersonic velocity through the hot and dilute CGM. A magnetic field is
assumed to lie parallel to the stream direction.
}
\label{fig:set-up}
\end{figure}

We study cold streams using both a 2D and a 3D model. The stream is a slab of
width $L$ in the 2D model while it is a cylinder of diameter $L$ (radius
$R=L/2$) in the 3D model, see Fig.~\ref{fig:set-up}.  Both models consider a
cold, dense stream moving at supersonic speed through the ambient hot, dilute
CGM. They thus act as idealized models of filamentary accretion of cold gas in
the CGM, a physical picture which is illustrated in fig. 1 in
\citet{Tumlinson2017}.

We take the stream density to be $\rhos = (1 + \delta)\rho_0$ where $\rho_0$
is the density of the CGM with $\delta>0$ for a dense stream. It is assumed
that the stream initially is in pressure equilibrium with the CGM which has
pressure $p_0$. This implies that the stream temperature is  $\Ts = T_0/(1 +
\delta)$ where $T_0$ is the CGM temperature. The stream is  thus colder and
denser than the CGM. As a consequence, the speed of  adiabatic sound waves in
the stream,  $c_{\mathrm{ad}, \mathrm{s}} = \sqrt{\gamma p_0/\rho_\mathrm{s}}$
where  $\gamma=5/3$ is the adiabatic index, is lower in the stream than in the
CGM  by a factor $1/\sqrt{1+\delta}$. The abrupt change in sound speed at the
surface of the stream means that it can act as waveguide for sound waves. Such
sound waves are called body modes because they permeate the entire body of the
stream and have an amplitude which decays exponentially with distance outside
the stream. Body modes are distinguished from surface waves  which are
localized at the surface of the stream and do not penetrate  far into it. This
distinction between surface and body modes is important for understanding the
physics of the supersonic KHI \citep{Payne1985,Hardee1988,Mandelker2016}.

The key difference between our models of cold streams and the ones studied by
\citet{Mandelker2016}, \citet{Padnos2018} and \citet{Mandelker2019} is that we
include a magnetic field and treat the system with ideal MHD (see e.g.
\citealt{Freidberg2014}). We consider a homogeneous magnetic field with
magnitude $B$ which is  oriented along the stream direction. The introduction
of a magnetic field  changes the wave properties of the fluid, i.e., sound
waves  are replaced by the compressive slow and fast magnetosonic waves and
the  shear Alfv\'{e}n wave. The properties of these MHD waves depend on
whether the propagation direction is parallel, perpendicular, or inclined with
respect to the direction of the magnetic field (see e.g. \citealt{Spruit2013}
for an introduction to MHD waves). The new characteristic wave speed
introduced by the magnetic field is the Alfv\'{e}n speed \be \va =
\f{B}{\sqrt{\mu_0\rho}} \ , \en where $\mu_0$ is the vacuum permeability. As
the Alfv\'{e}n speed depends on the gas density, it is lower in the stream
than in the CGM in our model (similarly to sound waves, by a factor
$1/\sqrt{1+\delta}$). The stream can therefore also act as a  waveguide in the
MHD limit. MHD waveguide models have a long history in modeling space plasmas
in the solar system (see e.g. \citealt{Mazur2010} and references  therein). A
key property of waveguides is that they have resonant modes, and  this leads
to resonances in the growth rate of the KHI.

The waves that arise in an MHD model of the stream can still be divided into
body modes, that have a large amplitude inside the stream, and surface modes,
which have  their amplitude localized at the surface of the stream. The
difference with  respect to hydrodynamics is that the waves are further
classified as  e.g. fast or slow body modes and fast and slow surface modes.
These classifications are carefully outlined in \citet{Edwin1982} for 2D slabs
and in \citet{Edwin1983} for 3D cylinders.  Their studies considered
\emph{stationary} systems but can still help us gain some intuition for the
dynamics of a \emph{moving} stream.

The stream moves through the CGM with supersonic speed, $V$, and  we define
the sonic Mach number of the stream velocity with respect to the
CGM\footnote{Because of the difference in sound speeds in the CGM and in the
stream, $\mathcal{M}_0$ differs from the Mach number defined with respect to
the stream itself, $\mathcal{M}_\mathrm{s} = V/c_{\mathrm{ad, s}}$, by
$\sqrt{\delta + 1}$.}, as \be \mathcal{M}_0 = \f{V}{c_{\mathrm{ad}, 0}} \ .
\en In our simplified model, a stream moving at speed $V$ through a stationary
medium, is physically equivalent to a \emph{stationary} stream  embedded in a
medium moving at speed $V$. This is a consequence of the system being
invariant under a Galilean transformation. Moving to this frame of  reference,
the stream can be regarded as a waveguide with moving boundaries. Body modes,
also known as reflective modes, are primarily localized inside of the stream
and gain energy by tapping into the free energy of the moving boundaries. This
can be explicitly shown by calculating the reflectivity and transmission
coefficients of waves as they impinge on the stream boundary. Such
calculations show that the amplitude of both transmitted and reflected waves
can be higher than the amplitude of the incident wave (\citealt{Payne1985} for
a 3D cylinder, \citealt{Hardee1988} for a slab). This energy gain at each
reflection gives rise to exponential  growth of wave amplitude, i.e., an
instability. Ingoing and outgoing waves can constructively interfere at
specific wavelengths, called resonant wavelengths, and the growth rate peaks
for these waves. Transmitted and reflected waves have nearly equal propagation
angles when such resonances occur and the resonant propagation angle can be
related to the Mach number of the stream (\citealt{Payne1985}, see also
\citealt{Mandelker2016}).

The waveguide analogy is not  perfect and only describes the linear dynamics
of the stream. In particular, the boundaries of the waveguide are not
immovable and not perfectly reflecting, and the surfaces of the stream will
eventually be disrupted by the large shear velocities present there. This
could occur in the form of the classical KHI  which is localized at the
surfaces.  For supersonic flow, however, a stabilization of surface modes
occurs \citep{Landau1944,Mandelker2016}. The instability that arises due to 
body modes can therefore be the dominant one when the flow is supersonic.

The key parameter, which  determines the importance of the magnetic field, is
the plasma-$\beta$
\be
    \beta = \f{p}{p_\mathrm{mag}} =2\f{c^2}{\va^2} \ ,
\en
where $p_\mathrm{mag} = B^2/(2\mu_0)$ is the magnetic pressure and
$c=\sqrt{p/\rho}$ is the isothermal sound speed. Since we assume an initially
constant pressure and magnetic field strength ($B=\mathrm{const.}$ and
$p=\mathrm{const.}$ in both stream and background), and $c$ and $\va$ have the
same density dependence, $\beta$ is also initially constant in space.

The idealized model outlined above of cold streams feeding  galaxies at high
redshift, contains three dimensionless parameters, $\rho_\mathrm{s}/\rho_0 = 1
+ \delta$, $\mathcal{M}_0$, and $\beta$. For a $10^{12} M_\odot$ dark matter
halo at redshift $z=2$, estimated values  for the density and Mach number of
the stream are $\rho_\mathrm{s}/\rho_0 \approx 10 - 100$ and $\mathcal{M}_0
\approx 0.75-2.25$ \citep{Mandelker2019}.  We take $\rho_\mathrm{s}/\rho_0=50$
and $\mathcal{M}_0=2$ throughout the main body of the paper and vary
only\footnote{A parameter study of growth rates for the 2D slab, where we vary
$\rho_\mathrm{s}/\rho_0$ and $\mathcal{M}_0$, is included in
Appendix~\ref{app:parameter-study}.} the magnetic field strength, i.e.,
$\beta$. We consider two different regimes for the magnetic field strength,
i.e.,  $\beta=1$, where the magnetic pressure is as large as the thermal
pressure, and $\beta=10$ where the magnetic pressure is 10 per cent of the
thermal pressure. These models are compared to a purely hydrodynamic reference
model ($\beta^{-1}=0$, no magnetic field).

The transition in values of density and velocity between stream and CGM have
so far been described as discontinuous. A discontinuous step function however
leads to growth rates that diverge with wavenumber which prohibits convergence
of numerical simulations
\citep{Robertson2010,McNally2012,Lecoanet2016,Berlok2019}. We therefore use a
smooth function, i.e., hyperbolic tangent profiles which have a smooth
transition on a length scale $\sim 2a$, where $a$ is the smoothing parameter.
This introduces a fourth dimensionless parameters, $a/L$. Note that this is
not just a numerical necessity, as a smooth profile is also physically
motivated due to turbulent mixing and diffusion. A high value of $a$ leads to
a significant reduction of the growth rate of the KHI while a low value makes
it difficult for simulations to converge \citep{Berlok2019,Mandelker2019}. We
choose a smoothing length $a/L=0.05$ which is a compromise between these two
extremes.

\section{Linear theory}
\label{sec:lin_theory}

We analyze the linear stability properties of the models by linearizing  the
ideal MHD equations in Cartesian and cylindrical geometry.  We use the linear
theory as an aid in understanding the results of  the \textsc{Athena++}
simulations that we present in Sections~\ref{sec:2D} and \ref{sec:3D}.  The
details of the linear stability analysis are presented in
Appendix~\ref{sec:linear_theory}, here we provide a brief outline of the steps
involved.

The key assumption for the linear theory is that the deviations from
equilibrium have the  form $f(z)\mathrm{e}^{-\mathrm{i} \omega t + \mathrm{i}
k x}$ for the 2D model  and $f(r)\mathrm{e}^{-\mathrm{i}\omega t + \mathrm{i}
k z + \mathrm{i} m \phi}$ for  the 3D model. Here, $\omega$ is a complex
frequency and  $\sigma = -\mathrm{Im}(\omega)>0$ is the growth rate of the
KHI. Linearizing the ideal MHD equations with this ansatz, one finds that the
resulting  equations constitute one-dimensional eigenvalue problems. We solve
these eigenvalue  problems by discretizing the equations with two different
types  of rational Chebyshev polynomials \citep{Boyd1987,Boyd1987a,Boyd} in
the  direction perpendicular to the stream. The procedure is performed with
\textsc{psecas}\footnote{\textsc{psecas}, Pseudo-Spectral  Eigenvalue
Calculator with an Automated Solver, is freely available online.}  which
automates some of the steps involved  \citep{Berlok2019}.

Given $\rho_\mathrm{s}/\rho_0 = 1 + \delta$, $\mathcal{M}_0$, $a/L$, $\beta$
and a wavenumber $k=2\upi/\lambda$ parallel to the stream (and for the
cylinder, an azimuthal mode number, $m$), \textsc{psecas} returns the
eigenvalues and eigenvectors of the eigenvalue problem. Here $\lambda$ is the
wavelength and the eigenvalues tells us how fast the KHI will grow. The
eigenvectors can be used to construct linear solutions for all the perturbed
variables, i.e., the density, temperature, magnetic field components etc.

\subsection{2D stream}
\begin{figure}
\includegraphics[trim = 0 15 0 10]{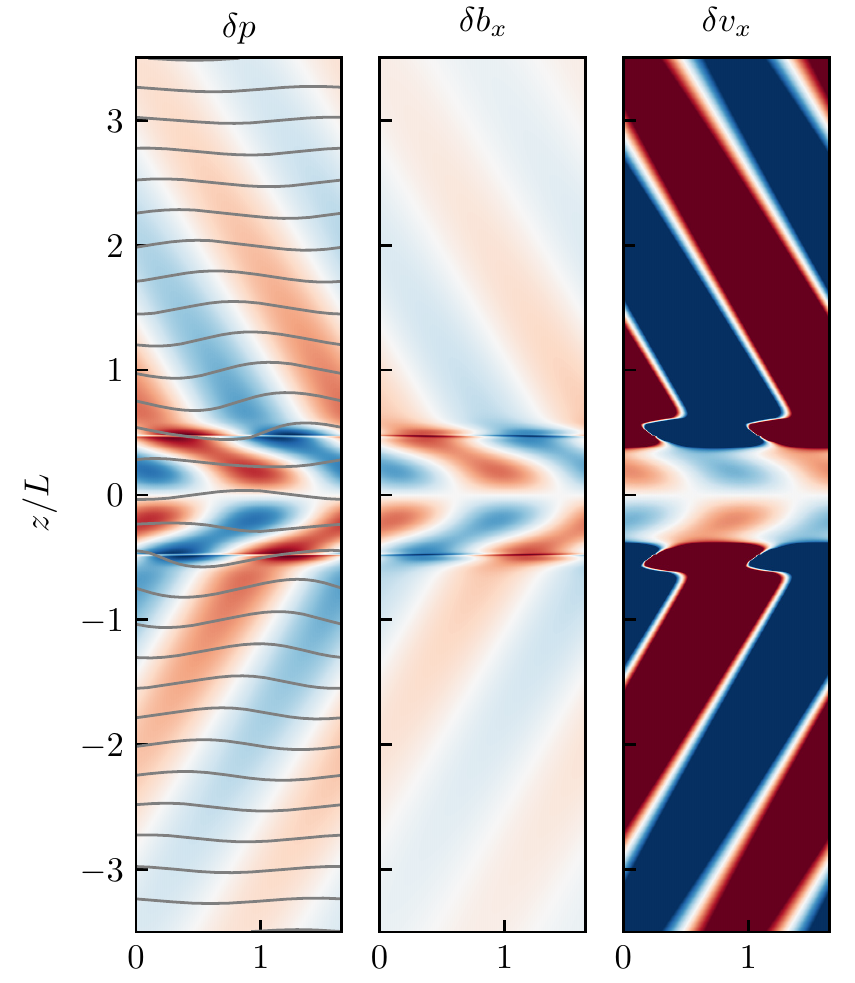}
\caption{The fastest growing eigenmode for the magnetized ($\beta=1)$,
supersonic ($\mathcal{M}_0 = 2$) KHI in Cartesian geometry. The (exaggerated)
perturbed  magnetic field lines (in the first panel) show that the mode is an
anti-symmetric sinusoidal mode. Inside the stream ($|z|<L/2$), the
perturbations are
magnetosonic body modes, which have pressure and magnetic  field, $\delta b_x$,
out of phase with the velocity, $\delta \varv_x$. Outside  the stream
($|z|>L/2$), the perturbations decay away from the interface.
}
\label{fig:eigmode-slab}
\end{figure}
We consider a 2D, dense and cold stream of gas moving at supersonic speed
through a hot and dilute background. Due to the abrupt change in velocity at
the surfaces of the stream, the stream is likely to be unstable to the KHI. We
are interested in understanding whether magnetic fields change this picture.
Do they stabilize the KHI? Do they change the non-linear evolution and prevent
the cold stream of gas from mixing with the CGM? Before introducing the extra
complexity of a magnetic field, we briefly outline how the KHI in a
supersonic, hydrodynamic stream already differs from the textbook picture of
the KHI that many readers are likely to have in mind.

For very high velocities, i.e., supersonic motions, the KHI changes it
behavior. In the classic textbook picture where a fluid  moving to the right
is separated from a fluid moving to the left (see e.g.
\citealt{Chandrasekhar1961}), the KHI arises at the interface between the two
fluids. If the motion is sufficiently supersonic, disturbances on the
interface are removed by the fast flow before they can grow. The precise
velocity at which this stabilization occurs depends on the density contrast,
$\delta$, and is given by equation 22 in \citet{Mandelker2016} who generalized
the result of \citet{Landau1944}.

In a supersonic stream, however, a different version of the KHI can arise in
which waves that are primarily localized inside the body of the stream grow in
amplitude as they are reflected back and forth between the inside surfaces. In
2D hydrodynamics, the body modes are the dominant type of instability in the
supersonic regime \citep{Mandelker2016}.

Let us now introduce a magnetic field, which is aligned with the stream. In
the classic textbook picture (with two separated fluids), this system has been
analyzed in \citet{Chandrasekhar1961} at subsonic velocities by considering
the incompressible limit of the MHD equations.  \citet{Chandrasekhar1961}
found that the KHI occurs at the interface of the two fluids (i.e., it is a
surface mode) and that the magnetic field stabilizes the KHI if the magnetic
field is sufficiently strong, that is if the Alfv\'{e}n speed is larger than
the gas velocity.\footnote{The same criterion also applies to the case of a
thin magnetic layer that separates the two counter-flowing fluids where the
magnetic layer can suppress instabilities on scales significantly larger than
its thickness \citep{Dursi2007}.}

The magnetized KHI for two fluids moving at supersonic speeds was studied in
\citet{Pu1983} who treated the problem using the full compressible equations.
As in \citet{Chandrasekhar1961} it was found that the flow velocity needs to
exceed the Alfv\'{e}n speed in order for the KHI to be unstable. Additionally,
it was found that highly supersonic flows stabilize the KHI and that the
solution found in this limit is given by stable, magnetosonic body modes.

In the hydrodynamic case, the supersonic stabilization only takes place when
there is a single interface between two fluids. When a 2D stream is considered
instead, sound waves are able to grow as body modes inside of it.  We find
that this picture carries over to the MHD version of the KHI. In particular,
we find that surface modes are stabilized by supersonic motion and magnetic
tension but that the KHI can instead take place as magnetosonic, body modes
that are primarily localized inside the stream. This is in agreement with
\citet{Hardee1992} who studied the magnetized slab in the context of an
astrophysical, under-dense jet.

We have calculated the growth rate of the KHI in the 2D stream as a function
of  wavenumber for the unmagnetized, $\beta=10$ and $\beta=1$ stream (see
Fig.~\ref{fig:theory-slab} in Appendix~\ref{app:2D-lin}).  In agreement with
\citet{Chandrasekhar1961}, we find that the magnetic  field inhibits the
growth rate of the KHI at low  wavenumbers where the modes are surface modes.
At higher wavenumbers, where the modes are body modes, the growth rate however
\emph{increases} with increasing magnetic field strength. We have the
following heuristic explanation for this surprising behavior. A strong
magnetic field can increase the restoring force of compressive  waves, i.e.,
as is well known for fast magnetosonic waves in a uniform medium. We believe
that the supersonic, magnetized KHI grows faster than the  unmagnetized
version, because the magnetic field in a similar way increases the  phase
velocity of reflective modes inside the stream.

We show the fastest growing eigenmode for $\beta=1$ in
Fig.~\ref{fig:eigmode-slab}. The magnetic field lines, shown in the first
panel with grey solid lines, are tied to the fluid in ideal MHD and thus trace
the fluid displacement. Opposite sides of the surface of the stream are
displaced in the same direction. This type of mode is called an anti-symmetric
sinusoidal mode. The mode has non-zero perturbations in pressure, density and
parallel magnetic field, which shows that the instability is compressive.

\subsection{3D stream}

\begin{figure}
\includegraphics[trim= 0 20 0 10]{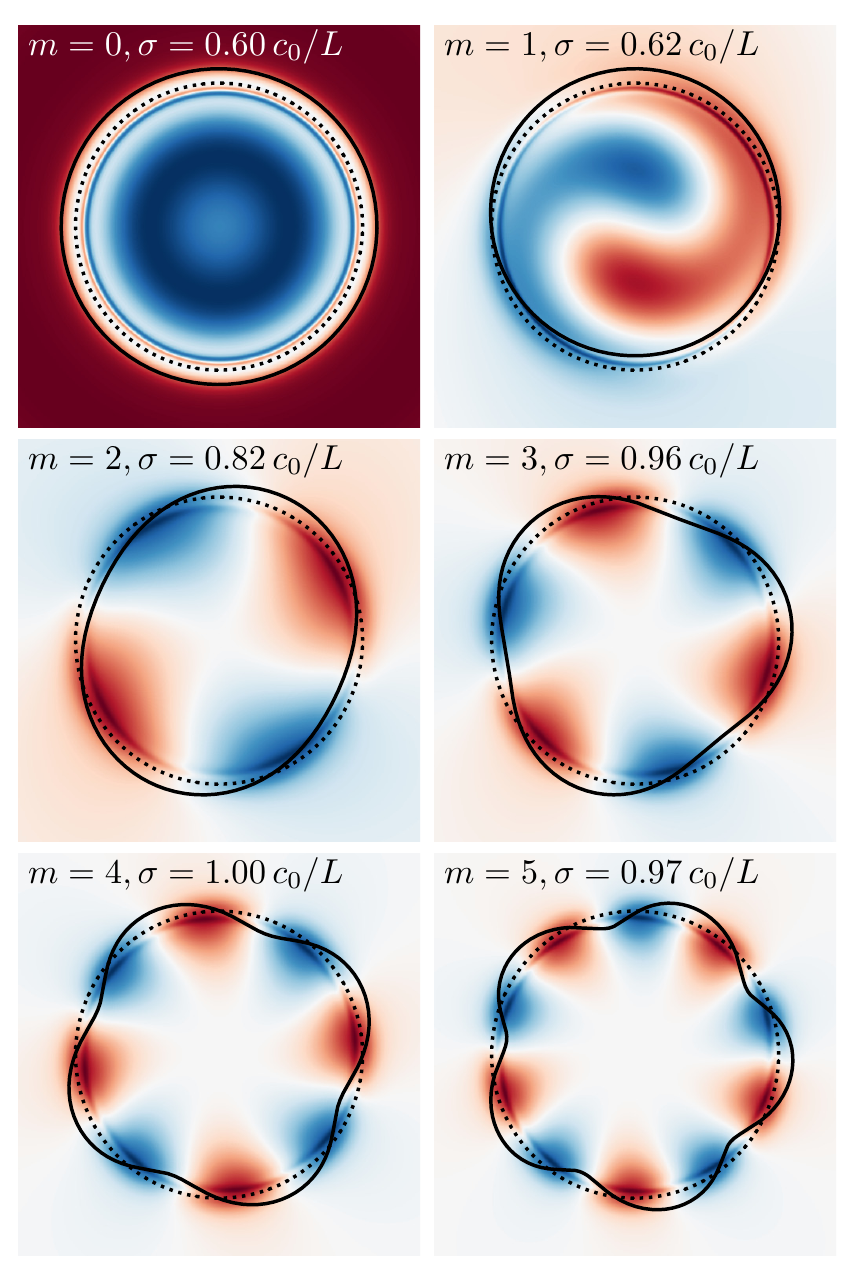}
\caption{Fastest growing eigenmodes for the magnetized
($\beta =1$), supersonic ($\mathcal{M}_0=2$) KHI in a cylindrical stream. We
show an $xy$-slice of the magnetic  field strength perturbation (which is in
phase with the pressure perturbation)  for the first 6 azimuthal modes
$m=0-5$.  The dotted circles indicate the initial interface between stream and
background and  the solid lines indicate how the shape is deformed  due to the
KHI.}
\label{fig:cylinder_modes}
\end{figure}

We analyze how the 3D cylindrical stream differs from the 2D, rectangular
stream.  The primary difference is, of course, the geometry and that the extra
dimension allows non-axisymmetric disturbances.  The linear theory for our 3D
cold stream model is closely related to the stability theory for astrophysical
jets, which can also be unstable to the KHI (e.g.
\citealt{Ferrari1981,Bodo1989,Appl1992,Bodo1996}). These studies analyze the
vortex sheet approximation, i.e., they consider a discontinuous transition
between a jet and the background, and derive analytical dispersion relations.
Our linear theory considers instead a smooth transition between stream and
background which requires a numerical treatment.

The eigenmodes are obtained with the aid of \textsc{psecas} and have the
general form $f(r)\mathrm{e}^{-\mathrm{i} \omega t + \mathrm{i} k z +
\mathrm{i} m \phi}$ where $f(r)$ is a radial dependence, $k$ is the wavenumber
along the stream and $m$ is the azimuthal wavenumber.  The azimuthal
dependence is illustrated in Fig. \ref{fig:cylinder_modes} where we show the
cross-section of the stream (an $xy$ slice at $z=0$) for the fastest growing
modes with $m=0-5$ and $\beta=1$.  These eigenmodes have been obtained with
\textsc{psecas} using linearized equations and a procedure which is detailed
in Appendix~\ref{app:3D-lin}. In Fig. \ref{fig:cylinder_modes}, the dotted
circle shows the initial cylinder ($r=R$) and the solid black lines how the
stream surface is displaced by the KHI (exaggerated here for illustrative
purposes). The underlying false color images show the perturbation to the
$z$-component of the magnetic field.

The $m=0$ mode is a body mode with disturbances throughout the interior of the
stream. In the MHD literature this mode is known as a sausage\footnote{The
deformed cylinder looks like a symmetrically squeezed sausage \citep[see e.g.
fig. 7 in][]{Fujimura2009}.} mode. The $m=1$ mode is the 3D equivalent of the
sinusoidal mode that we found to be the fastest growing mode for the 2D
stream, see Fig.~\ref{fig:eigmode-slab}. This mode is also known as a kink
mode in the MHD literature \citep[e.g.][]{Fujimura2009}.

For higher $m$ (called fluting modes), the instability appears as a mixture of
surface and body modes, with a pronounced disturbance at the surface of the
cylinder and disturbances penetrating gradually less into the cylinder as $m$
increases. That is, the instability gradually changes from being dominantly of
the body mode type to being dominantly a surface mode as $m$ increases.  These
surface modes survive stabilization by the supersonic flow because the
variation in $z$ has long wavelength, making the disturbance effectively
subsonic\footnote{Surface modes are suppressed when a supersonic flow removes
a perturbation faster than it grows \citep{Mandelker2016}. This process is
faster on short spatial scales than on large spatial scales.
\citet{Mandelker2016} shows explicitly how so-called fundamental modes are
unstable at long wavelength for any value of $\mathcal{M}_0$ (and $\delta$,
see their Sections 2.3.2 and 2.4.1 for a 2D slab and 3D cylinder,
respectively).}, and because the azimuthal variation is perpendicular to the
flow and thus unaffected by magnetic tension.

The growth rate of the KHI in 3D depends on the additional parameter, the
azimuthal wavenumber, $m$. We have calculated the growth rates as a function
of $kL$ and $m$ in Appendix \ref{app:3D-lin}, see
Fig.~\ref{fig:theory-cylinder}. This allowed us to find the value of $kL$
which gives the fastest growth for each azimuthal wavenumber, $m$. The fastest
growing mode is an $m=4$ surface mode in both the unmagnetized  and the
magnetized limit.

\section{MHD Simulations}
\label{sec:simulations}

\begin{figure*}
\includegraphics[trim = 0 20 0 10]{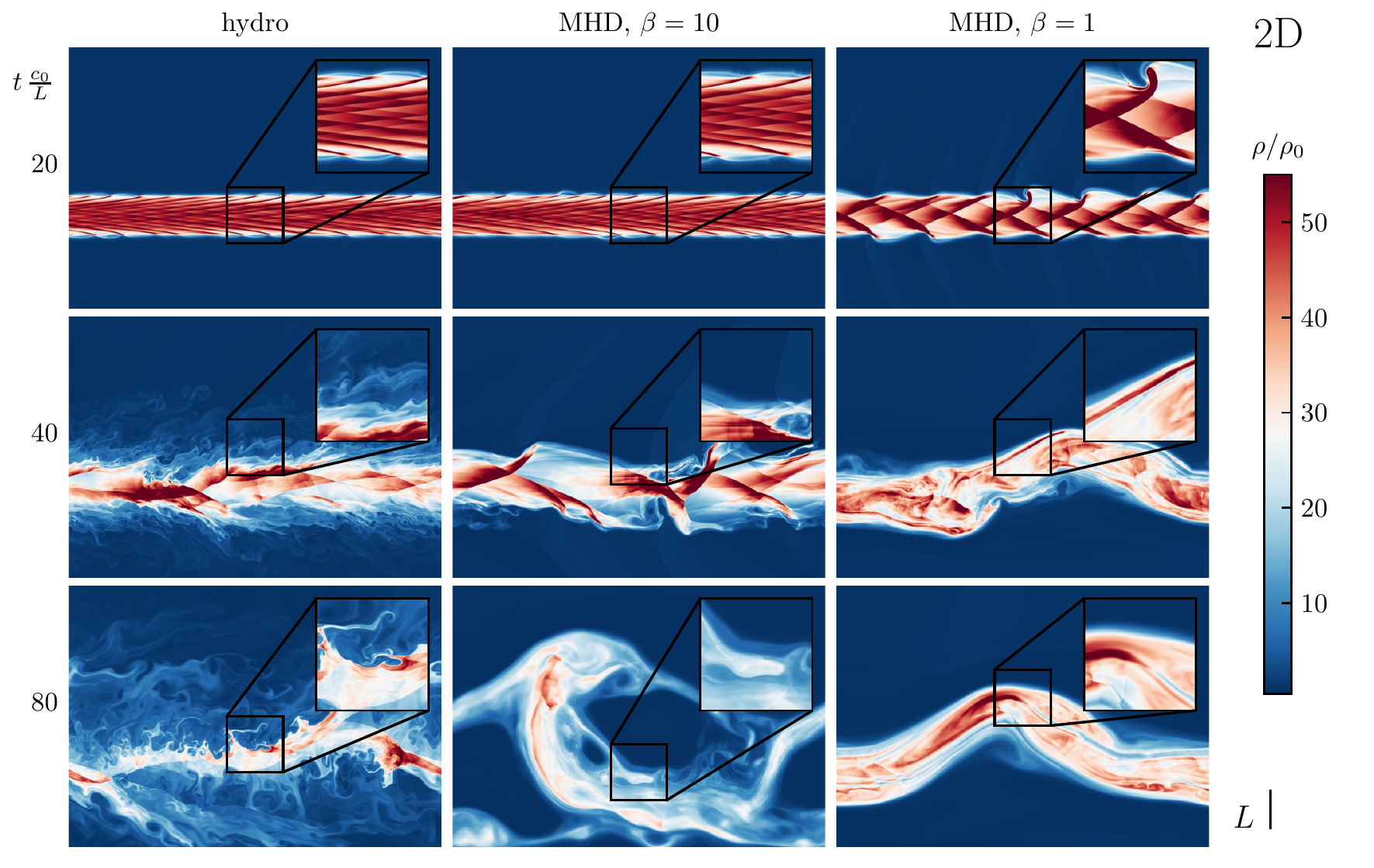}
\caption{ 2D stream: density snapshots of the disruption of a cold, dense stream
  of width $L$ propagating with supersonic speed through the hotter, less dense
  CGM. Zoom insets have size $3/2L\times 3/2L$ and highlight small scale mixing (or
  lack thereof).  Magnetic fields dramatically reduce mixing of the stream with
  the CGM.  }
\label{fig:2D_snaps}
\includegraphics[trim = 0 15 0  0,clip=True]
{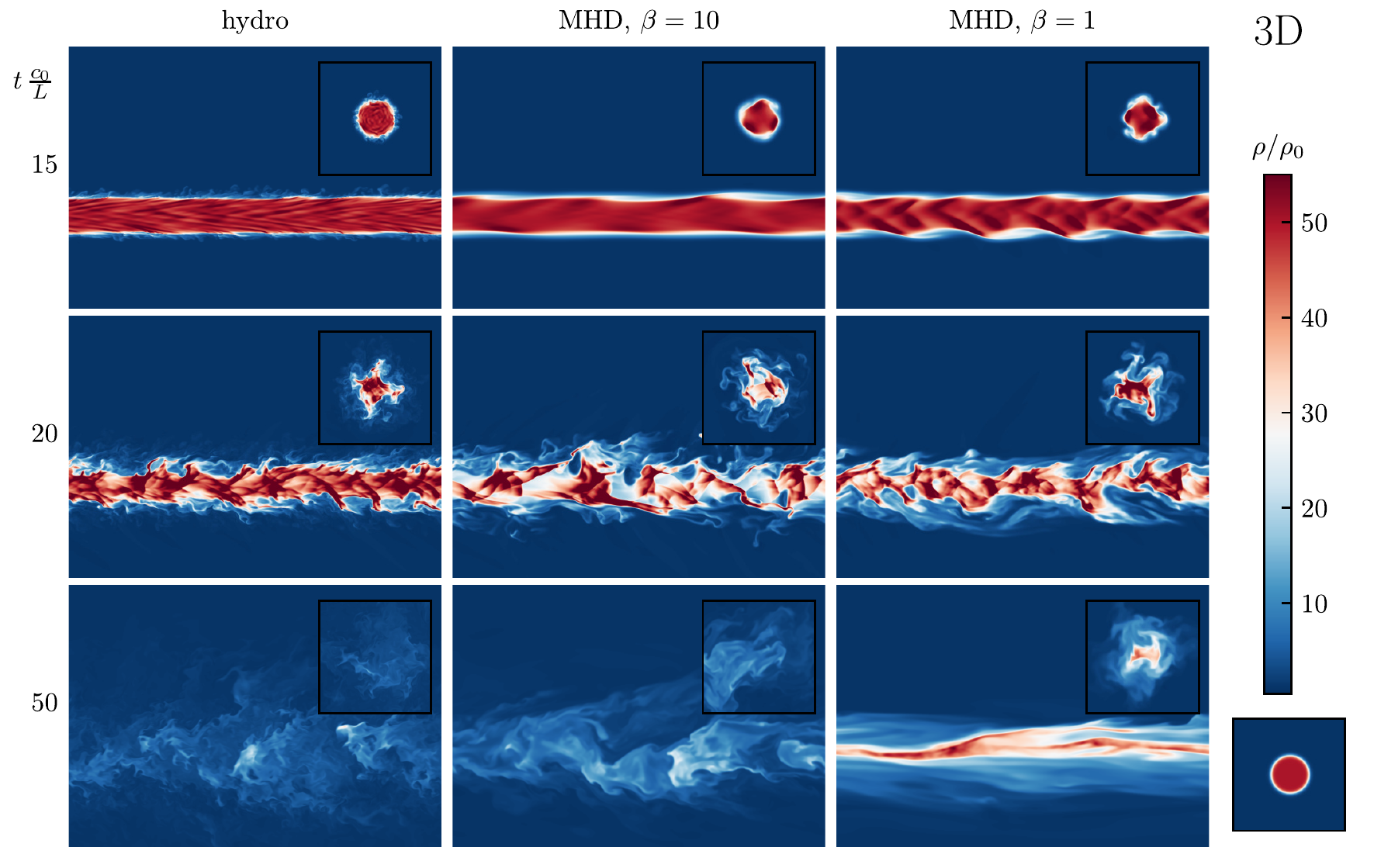}
\caption{ 3D stream: density snapshots of cylindrical cold dense streams. Main
  panels show the stream from the side and insets show a cross-section.
  Magnetic fields still suppress mixing in 3D, although not as dramatically as
  in 2D.}
\label{fig:3D_snaps}
\end{figure*}

The computer simulations are performed with the MHD code \textsc{Athena++}
\citep{White2016} on a Cartesian domain. We perform three simulations in
each geometry (2D and 3D), i.e., a pure hydrodynamic simulation and MHD
simulations with $\beta=10$ and $\beta=1$, for a total of 6 simulations.

The 2D simulations have size $L_x \times L_z = (10\times200)L$ and the 3D
simulations have size $L_x \times L_y \times L_z = (200\times200\times10)L$
where $L$ is the width of the stream. The stream is moving in the
$x$-direction in the 2D simulations and in the $z$-direction in the 3D
simulations. The long domain size along the stream is chosen in order to
resolve high wavelength modes which can have significant growth rates (see
Appendix \ref{sec:linear_theory}). The highly elongated domain in the
perpendicular direction(s) is chosen in order to mimic an infinite domain. We
employ periodic boundary conditions in the parallel direction and outflow
boundary conditions in the perpendicular direction(s).

The interesting stream dynamics and disruption take place near the initial
position of the stream, and covering the entire domain with uniform resolution
would be wasteful and computationally unfeasible.  We therefore enable the
static mesh refinement capabilities of \textsc{Athena++} with 10 levels of
refinement in the 2D simulations and 8 levels of refinement in 3D simulations.
This increases the resolution by a factor of two at each level of refinement.
The 2D simulations have $\Dx=\Delta z= 10\times2^{-12} L \approx 2.4\times
10^{-3}L$  inside the central $2L$ while the 3D simulations are four times
coarser.  This corresponds to $\sim 400$ ($\sim 100$) cells per stream width
in 2D (3D).

The simulations are initialized with subsonic, random Gaussian velocity noise,
with a standard deviation of $10^{-3} c_0$ in each (Cartesian) component.  We
seed every cell in the computational domain and do not restrict the
disturbances to the surface of the stream.  All simulations employed the
second order accurate Van Leer scheme with a Courant number of 0.4 (0.3) for
the 2D (3D) simulations and the HLLE (HLLC) Riemann solver for the MHD
(hydrodynamic) simulations.\footnote{HLLE is more diffusive than the HLLD
Riemann solver but the 2D simulations with $\beta=1$ crash when we use HLLD.
We believe this happens due to the carbuncle instability\citep{Quirk1994}, a
numerical instability that can arise with grid-aligned shocks (see discussion
in Appendix C in \citealt{Stone2008}).}

The simulations are performed in a frame in which the stream is moving at
speed $V/2$ and the background (the CGM) is moving in the opposite direction
at the same speed. This is of course physically equivalent to a stream moving
at speed $V$ through a stationary medium. Numerically, however, the change of
reference frame reduces the speed with respect to the grid, which reduces
advection errors and increases the allowed time step. All analysis of
simulations is performed in the natural frame, i.e., a moving stream and
stationary CGM.\footnote{Note that the dense stream in our MHD simulations
moves to the left, unlike our linear theory analysis, where it moves to the
right. The physics results are, of course, invariant under this change of
frame.}

\subsection{2D stream}
\label{sec:2D}

The initial condition for the stream velocity and density in the simulations
is
\begin{align}
    \varv(z) &=  \f{V}{2} \left[\tanh\left(\f{z+L/2}{a}\right) -
                       \tanh\left(\f{z-L/2}{a}\right) - 1 \right]\ ,
    \label{eq:v-equi} \\
    \f{\rho(z)}{\rho_0} &=
    1 + \f{\delta}{2} \left[\tanh\left(\f{z+L/2}{a}\right) -
                       \tanh\left(\f{z-L/2}{a}\right)\right] \ ,
    \label{eq:rho-equi}
\end{align}
and the magnetic field is aligned with the stream velocity, $\vec{B}=B\ex$.

Fig.~\ref{fig:2D_snaps} compares the density evolution of the stream in a
hydrodynamic simulation (left-hand column) with MHD simulations with $\beta =
10$ (middle column) and $\beta = 1$ (right-hand column). Density disturbances
due to the KHI arising as body modes are seen in the top row of panels. The
criss-cross pattern inside the streams at $t=20 L/c_0$ in the hydrodynamic and
$\beta=10$ simulations are signatures of the spectrum of body modes that grow
inside of them (see insets). At this point in time, the simulations are about
to enter the nonlinear stage in which weak oblique shocks form
\citep{Padnos2018}. The criss-cross pattern arises already earlier in the
linear regime, which distinguishes the pattern as a superposition of body
modes that are constructively interfering.  The case of $\beta=1$, which has a
larger growth rate, is already in the nonlinear regime at $t=20 L/c_0$, and
the interior of the stream is being disrupted by a series of overlapping weak
oblique shock and rarefaction waves.  The shock waves propagate from the
stream into the background medium at an angle and transfer parallel kinetic
energy and momentum from stream to the background. Due to momentum
conservation, an oblique rarefaction wave is generated at the boundary that
propagates back into the stream. This leads to a deceleration of the stream
which we discuss in more detail in Section~\ref{sec:2D-vs-3D}.

Quite remarkably, the cold stream starts to disrupt on a faster time scale in
the MHD simulation with $\beta=1$ than in hydrodynamic simulation.  This is a
surprising result as magnetic fields inhibit the KHI in the incompressible,
subsonic regime \citep{Chandrasekhar1961}.  The faster time scale for growth
of the strongly magnetized KHI is predicted by the linear theory in
Section~\ref{sec:lin_theory}. We believe that the  increased growth rate
occurs as a consequence of an increased propagation speed of waves.
Physically, an increase in phase speed allows more reflections per unit time.
This could explain the increased growth rate, assuming that the energy gain
per reflection is roughly constant. This interpretation, is somewhat supported
by analytical studies of hydrodynamic body modes where \citet{Mandelker2016}
derived that the stability criterion and growth rate for body modes depends on
the sum of the sound speed inside and outside of the stream \citep[see
equations 32 and 36 in][]{Mandelker2016}. Precisely how this works in the
magnetized case is left for future studies.

The middle row of panels in Fig.~\ref{fig:2D_snaps} show the streams at
$t=40L/c_0$. The streams all have clear signatures of shock waves,
with sharp density discontinuities that overlap and form complex
structures. The strongly magnetized simulation $\beta=1$ has acquired a bend
but has otherwise retained its structural integrity. This is in contrast to
the hydrodynamic and $\beta=10$ streams which have started to disrupt.

As evident from Fig.~\ref{fig:2D_snaps}, the mode that ends up disrupting the
stream is not necessarily the fastest growing one. The reason is two-fold.
Firstly, because we excite the instability with white noise and a finite
amplitude, slower growing modes might be the first to disrupt the stream if
their initial amplitude is large (as this gives them a head start). Secondly,
as pointed out in \citet{Padnos2018,Mandelker2019}, the fastest growing mode
might saturate and cease exponential growth without breaking up the stream.
This happens in the $\beta=1$ simulation where the disturbance seen at
$t=20L/c_0$ roughly corresponds to the fastest growing mode but the bending at
$t=80L/c_0$ occurs at a much longer wavelength. Physically, the magnetic field
opposes bending of the stream because bent magnetic field lines have an
associated magnetic tension. This magnetic tension is larger on small scales
than on large scales (the tension force scales with $k^2$, see
Equation~\ref{eq:mom-z-lin}). Magnetic tension therefore prevents small
wavelength bending of the stream while still allowing large scale bending (for
this magnetic field strength).

The small scale mixing of stream material with the CGM is also suppressed by
magnetic fields. This is most clearly seen in the insets in the lower row of
panels (at $t=80L/c_0$). In the MHD simulations the boundary between stream
material and CGM remains sharp and clearly defined. In contrast, the boundary
in the hydrodynamic simulation is less clearly defined. Instead, wispy
filaments of stream material are mixed into the CGM on small scales. The
difference in the mixing properties is due to magnetic tension, which becomes
important on small scales.

Based solely on the 2D simulations, it seems that magnetic fields could very
significantly modify cold stream disruption and mixing, even for a
sub-dominant magnetic field ($\beta=10$). We will see that such a strong
conclusion cannot be made in 3D, where the extra degree of freedom enhances
the mixing rate in both hydrodynamic and MHD simulations. A quantitative
analysis of stream mixing in 2D is therefore postponed to
Section~\ref{sec:2D-vs-3D} such that we can compare with the 3D simulations.

\subsection{3D stream}

\label{sec:3D}

We extend the study of a dense and cold stream of gas moving with supersonic
speed through a hot, dilute background to 3D. The stream
has a cylindrical shape with diameter $L$ (radius
$R = L/2$) and is cold and dense inside $r<R$. We again assume a smooth
transition of velocity and density now given by
\begin{align}
    \varv(r) &= \f{V}{2} \tanh\left(\f{r-R}{a}\right) \ ,
    \label{eq:v-equi-cyl} \\
    \rho(r) &= \rho_0 + \rho_0 \f{\delta}{2}\left[ 1 - \tanh\left(\f{r-R}
    {a}\right)\right] \ ,
    \label{eq:rho-equi-cyl}
\end{align}
where $a$ is the smoothing length.
The stream moves in the $z$-direction, $\vec{\varv} = \varv(r)\ez$ and
the stream direction is aligned with a constant magnetic field, $\vec{B}=B\ez$.

We show density snapshots from the three (hydro, $\beta=10$ and $\beta=1$) 3D
simulations in Fig.~\ref{fig:3D_snaps}. The main panels show cuts down the
middle of the stream ($xz$-slices with $y=0$) with the perpendicular extent
limited to the interesting $x\in[-2.5, 4.5]L$. The insets show cross-sections
of the stream ($xy$-slices at $z=0$). The cross-section at $t=0$ is shown
beneath the color bar on the same scale as the rest of the panels.

The first row of panels compares the three simulations at $t=15 L/c_0$. The
disturbances in density take place at the surfaces of the streams. This
differs from the 2D simulations where the disturbances arose as body modes
inside the stream. We can understand this difference between
2D and 3D simulations by using results from linear theory.
The body modes were dominant in the 2D simulations because surface modes with
surface variation along the stream direction are suppressed by supersonic
flows. In 3D, surface modes with azimuthal surface variation are however not
suppressed, and are predicted to grow faster than the body modes (see
Figs.~\ref{fig:cylinder_modes} and \ref{fig:theory-cylinder}).

The 3D simulations have much faster growth and mixing than the 2D simulations.
This is evident in the second row of panels where the streams have begun their
disintegration at $t=20 L/c_0$. The boundary between
stream and CGM is much less clearly defined than in the corresponding 2D
simulations, even for the $\beta=1$ simulation.
The key difference is that the 3D simulations allow azimuthal motions that
are unaffected by magnetic tension. As a result, the lower row of panels,
at $t = 50 L/c_0$, reveals that
even the $\beta=1$ is eventually disrupted and mixed into the CGM. We compare
the quantitative differences in stream mixing in the following section.

\subsection{Comparison between 2D and 3D}
\label{sec:2D-vs-3D}

\begin{figure*}
\includegraphics[trim = 0 20 0 10]{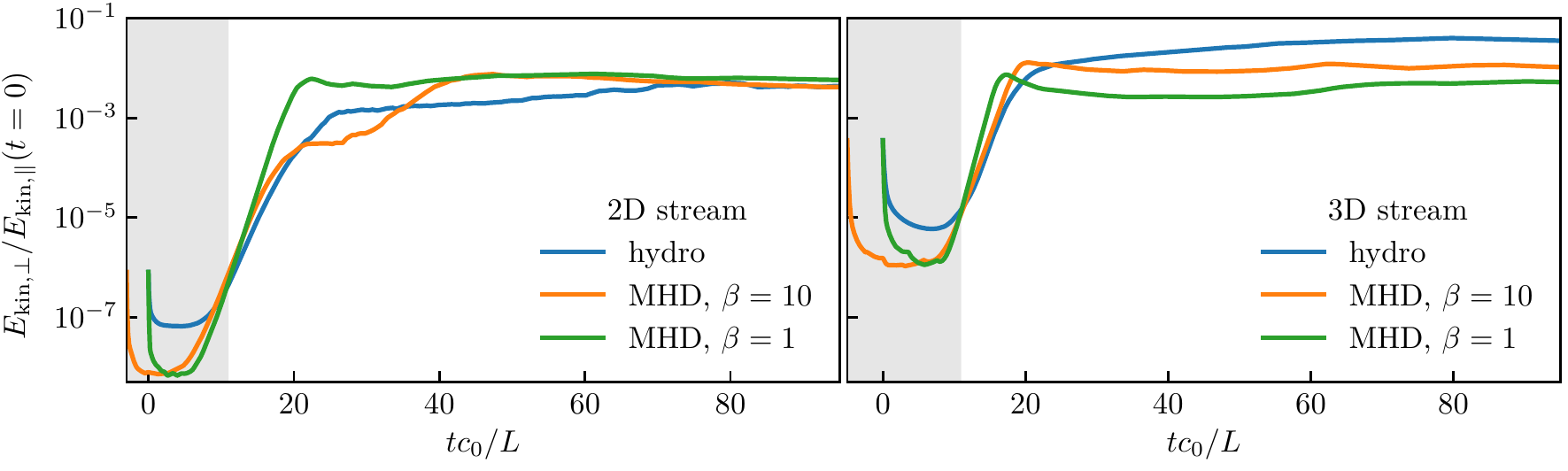}
\caption{Growth of total kinetic perpendicular energy, $E_{\mathrm{kin}, \perp}$,
relative to the initial parallel kinetic energy, $E_{\mathrm{kin}, \para}(t=0)$.}
\label{fig:perp_energy}
\includegraphics[trim = 0 20 0 0]{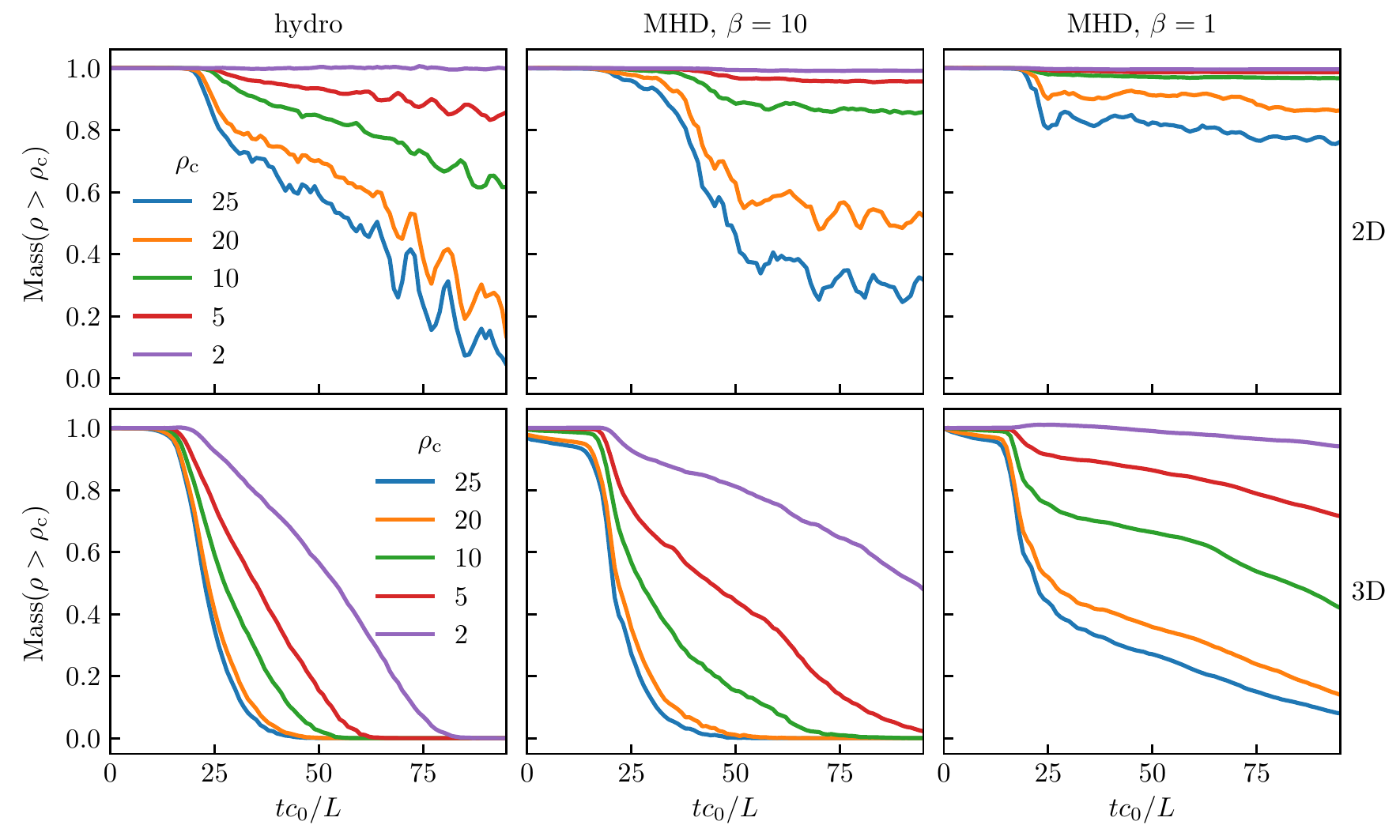}
\caption{Time evolution of the sum of mass with a density above a certain
  cutoff, $\rho_{\mathrm{c}}$. The 2D (3D) simulations are shown in the upper
  (lower) panel for five different cutoffs, $\rho_\mathrm{c}/\rho_0 = 25, 20,
  10, 5$ and 2. In 2D, magnetic fields strongly suppress mixing, independently
  of the value of $\rho_\mathrm{c}$ considered. This effect is less dramatic in
  3D, but magnetic fields with $\beta=1$ still significantly slow the rate of
  mixing.  For the 3D simulation with $\beta=10$, only mixing to densities
  $\lesssim 2-5 \rho_0$ is signicantly slowed down, while mixing to densities
  $\lesssim 10-25 \rho_0$ occurs at almost the same rate as in the 3D
  hydrodynamic simulation.  }
\label{fig:dense-fraction-detailed}
\end{figure*}

In all cases, some
form of the KHI
(i.e., arising as surface or body modes, magnetized or unmagnetized) leads to
exponential growth of kinetic energy in the perpendicular direction. The
streams initially have total, parallel kinetic energy
\be
    E_\mathrm{kin,\para} = L_x \int_{-\infty}^\infty \frac{\rho(z)}{2} \varv(z)^2
    \,
    dz
\en
in 2D and
\be
    E_\mathrm{kin,\para} = 2\upi L_z \int_0^\infty \frac{\rho(r)}{2} \varv(r)^2 r \,
    dr
\en
in 3D. We show the time evolution of the perpendicular kinetic energy,
relative to this initial parallel kinetic energy, in
Fig.~\ref{fig:perp_energy}. For both 2D and 3D simulations, the initial
exponential growth is faster in the $\beta=1$ simulations than in the
hydrodynamic simulations, as expected from the linear theory.

We seed the simulations with random Gaussian velocity components, which
initially decay before growing exponentially. This initial transient in
perpendicular kinetic energy, seen at $t\lesssim 10 L/c_0$, differs between
hydrodynamic and MHD simulations, with a more severe reduction in the initial
energy in the MHD simulations. This is because a fraction of the kinetic
energy of velocity fluctuations is shared with the magnetic field on all
length scales via the induction equation, reducing the initial kinetic energy
more severely in comparison to the hydrodynamic case, which can dissipate this
energy only on small scales.  The $\beta=10$ simulation therefore displays a
delay in the evolution.\footnote{The $\beta=1$ KHI has a larger growth rate
than the hydrodynamic KHI and the $\beta=1$ simulation quickly catches up
despite the larger initial decay. We explore an alternative seeding mechanism
in Appendix~\ref{sec:verification} where we use the fastest growing eigenmode
to initialize simulations.} In order to compare the different simulations on
an equal footing, we have shifted the definition of $t=0$ by $3 L/c_0$ ($5
L/c_0$) for the 2D (3D) $\beta=10$ simulation in all analysis and figures.
These shifts cause the simulations to approximately exhibit the same
perpendicular kinetic energy at $t=10L/c_0$.

We have seen that the stream disintegrates faster in 3D than in 2D  because
the extra degree of freedom allows non-axisymmetric fluting modes  with $m>0$
to destabilize the stream. Visual inspection of Figs.~\ref{fig:2D_snaps} and
\ref{fig:3D_snaps} also shows that the mixing occurs much faster in 3D and
that the suppression of mixing by magnetic fields is less efficient. We now
perform a more quantitative analysis in order to understand the consequences
for the feeding  of cold gas to massive galaxies at high redshift.

The stream initially has density $\rho_\mathrm{s} = 50 \rho_0$.  A stream
fully mixed with the CGM would give a uniform medium with density $\approx
1.24 \rho_0$ ($\approx 1.001\rho_0$) in the 2D (3D) simulations. We study
stream mixing with an observationally motivated criterion and calculate the
time evolution of the total mass of gas with density above a certain cut-off,
$\rho_\mathrm{c}$. This quantity is shown in
Fig.~\ref{fig:dense-fraction-detailed} for various values of the cut-off,
i.e., $\rho_\mathrm{c}/\rho_0=25, 20, 10, 5$ and 2. The values have been
normalized to their value at $t=0$, for easier comparison between 2D and 3D.

The 2D results (upper row in Fig.~\ref{fig:dense-fraction-detailed}) show
clear suppression of mixing by magnetic field, with gas with $\rho>10\rho_0$
totaling 97 per cent of the mass of the initial stream mass at the end of the
$\beta=1$ simulation.  With $\beta=10$, this number is 86 per cent, which is
still significantly higher than the 57 per cent found for the hydrodynamic
simulation.

The 3D simulations have more mixing in general and less suppression of mixing
by the magnetic field (lower row in Fig.~\ref{fig:dense-fraction-detailed}).
Here we find that gas with density $\rho>10\rho_0$ totals 39 per cent of the
initial stream mass at the end of the $\beta=1$ simulation but that this
number is 0 per cent in both $\beta=10$ and hydrodynamic simulations. And
although the mixing of material with $\rho>10\rho_0$ occurs at a slightly
slower rate in the $\beta=10$ than in the hydrodynamic simulation, this
difference does not seem significant.

There is however a non-trivial dependence on the value of $\rho_\mathrm{c}$
considered. For instance, the difference between the hydrodynamic and  the
$\beta=10$ simulation is quite significant when $\rho_\mathrm{c}/\rho_0=2$  or
5 is  considered. In the $\beta=10$ simulation, gas with $\rho>2\rho_0$ still
totals more than 50 per cent of the initial stream mass at the end of
simulation while this fraction goes to zero already at $t \sim 80 L/c_0$ in
the hydrodynamic simulation.

This survival time criterion is based on the amount of mass that can retain
densities above a certain cut-off, $\rho_\mathrm{c}$. It is also important to
assess whether the stream material can remain cold. Mixing of stream with CGM
that reaches a mean density $\rho_\mathrm{c}$ yields a mean temperature
$T_\mathrm{c} = T_0 \rho_0/\rho_\mathrm{c}$ (if we assume perfect mixing of
the internal energy content and neglect heating/cooling mechanisms). The
mixing temperature is lower than the CGM temperature but higher than the
initial stream temperature by a factor $\rhos/\rho_\mathrm{c}$. A cutoff of
$\rho_\mathrm{c}/\rho_0 = 5$ thus also corresponds to a factor 10 increase in
temperature if the internal energy is mixed. Our simulations are unfortunately
not able to capture this important aspect of the evolution as cooling and heat
conduction is not included in the analysis. An important future extension of
our work is therefore to include radiative cooling in order to understand the
thermal history of the cold stream. The combination of magnetic fields and
radiative cooling has been shown to increase the efficacy of thermal
instability \citep{Suoqing2018}.

\section{Astrophysical implications}
\label{sec:astro-impli}

\begin{figure}
\includegraphics[trim = 0 20 0 0]{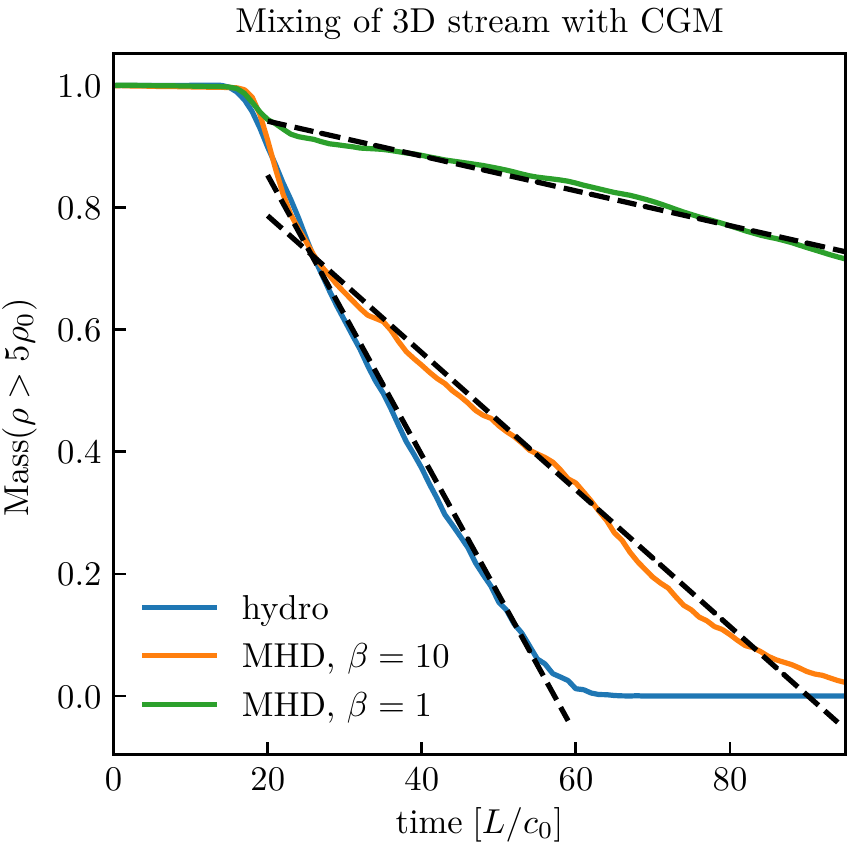}
\caption{Mixing of cold dense streams with the hot, dilute CGM. The mixing
rate is $\sim 2$ times lower in MHD simulations with $\beta=10$ than in a
reference hydrodynamic simulation. For $\beta=1$, the mixing rate is
suppressed by a factor $\sim 8$.}
\label{fig:disruption_sum}
\end{figure}

We show the amount of mass with $\rho>5\rho_0$ for the 3D simulations in
Fig.~\ref{fig:disruption_sum}. As a result of the non-linear phase of the KHI
the dense stream disintegrates and eventually mixes with the ambient CGM,
lowering the mass of the dense stream. For this to happen, the conversion of
parallel to perpendicular kinetic energy, $E_{\mathrm{kin}, \perp}$, must have
saturated. Consequently, while $E_{\mathrm{kin}, \perp}$ is still growing
($t\lesssim 20 L/c_0$) there is no mixing observed in either model.  We
observe that mixing of cold stream gas with the CGM occurs at an approximately
constant rate for $t\gtrsim 20 L/c_0$ but that the rate at which it proceeds
depends on the magnetic field strength.

We perform a linear fit, indicated with black dashed lines, in order to
estimate the mixing rates. We find a mixing rate of $-0.023\,c_0/L$ for the
hydrodynamic simulation, which is $\sim 2$ higher than the one found in the
$\beta=10$ simulation ($-0.011 \,c_0/L$), and $\sim 8$ times higher than the
$\beta=1$ simulations ($-0.003\,c_0/L$). Magnetic fields can thus inhibit the
mixing of cold streams with the CGM.  We can convert the mixing rates to
half-live survival times, i.e., how long it takes mixing to cause less than
half the mass to have $\rho>5\rho_0$.  We find half-life survival times of
$t_{1/2} \sim 20 L/c_0$, $t_{1/2} \sim 45 L/c_0$ and $t_{1/2} \sim 175 L/c_0$,
for the hydrodynamic, $\beta=10$ and $\beta=1$ simulations, respectively. The
decreased mixing rates thus leads to increased survival times of magnetized
streams.

This enables us to convert the estimated survival times into an estimated
lower limit on the width of cold streams that can make it to the central
galaxy.  The cold stream in our model has $\mathcal{M}_0 = 2$, i.e., velocity
$V = M \sqrt{\gamma} c_0$, such that the distance traveled by the stream
during its half-life is
\be
    V t_{1/2} = \xi M \sqrt{\gamma} L \ ,
\en
with $\xi \approx 20$, 45 and 175 for the hydrodynamic, $\beta=10$ and
$\beta=1$ simulations, respectively. Equating this estimate for the traversed
distance with the virial radius of the halo, we find that streams with
\be
    \f{L}{R_\mathrm{v}} \lesssim \f{1}{\xi M \sqrt{\gamma}}
    =
    \left\{
    \begin{array}{lll}
    2 \times 10^{-2} &\mathrm{for} & \mathrm{hydro} \\
    9 \times 10^{-3} &\mathrm{for} & \beta=10 \\
    2 \times 10^{-4} &\mathrm{for} &\beta = 1 \\
    \end{array}
    \right.
     \ ,
\en
will have less than half their mass at densities $\rho>5\rho_0$ when they
reach the central galaxy. In other words, cold streams thinner than this limit
will be disintegrated by the KHI before they can reach the central galaxy. The
decreased mixing rate of cold streams with the CGM can thus make it possible
for thinner streams to reach the central galaxy when magnetic fields are
included. We find that a magnetic field with $\beta=10$ has a significant
effect while $\beta=1$ has a dramatic effect. We now estimate the required
magnetic field strength in physical units.

The magnetic field strength is related to the hydrogen
number density, $n_{\mathrm{H}}$, temperature of the hot, dilute CGM gas,
and the plasma-$\beta$ by
\be
    B = \sqrt{\f{2p\mu_0}{\beta}} =
    0.8 \left(\f{n_{\mathrm{H}}}{10^{-4} \mathrm{cm}^{-3}}\right)^{1/2}
    \left(\f{T}{10^{6} \mathrm{K}}\right)^{1/2}
    \beta^{-1/2} \mu\mathrm{G} \ .
    \label{eq:B_in_muG}
\en
For $n_{\mathrm{H}}=10^{-4} \mathrm{cm}^{-3}$ and $T=10^{6}$ K, characteristic
values for the CGM \citep{Tumlinson2017}, Equation~\eqref{eq:B_in_muG} yields
$B\approx 0.8 \mu\mathrm{G}$ for $\beta=1$ and $B\approx 0.25 \mu\mathrm{G}$
for $\beta=10$. Magnetic field strengths larger than $0.25 \mu$G are therefore
predicted to modify the mixing rate of cold streams with the CGM, their
survival times, and consequently the minimum widths of streams which can make
it to the central galaxy.

Whether such magnetic field strengths are found in filaments or the CGM is
currently unknown (see also the discussion in the introduction).  The magnetic
fields present in the CGM are thought to be generated in the galactic disk and
subsequently transported into the CGM \citep{Marinacci2018, Nelson2019}.
Observations of radio halos show equipartition magnetic fields of order
7--10~$\umu$G at distances of 6~kpc from the galaxies' midplane
\citep{Stein2019}. While these values are in line with synchrotron modelling
of advection-dominated, accelerating galactic winds \citep{Miskolczi2019},
advection-dominated models with a constant wind speed yield values that are
smaller by a factor of 2.5 \citep{Heesen2018}. Cosmological simulations of
galaxy formation predict magnetic field strengths in the required range at a
radial distance of 30~kpc in a Milky Way-like disk galaxy at $z=2$ (see fig.~6
in \citealt{Pakmor2014}).

The introduction of magnetic fields in our models also required setting a
topology. As for the magnetic field strength, to date this cannot be
observationally inferred.  We have assumed a simple configuration with a
uniform magnetic field of constant field strength aligned with the stream both
inside and outside of the stream.  This choice is supported by cosmological
simulations that found that magnetic fields are aligned with filaments in the
cosmic web \citep{Brueggen2005}. The choice is not well motivated for the CGM
in which the magnetic field is likely to be turbulent and tangled.  Whether
our results are robust to changes in these assumptions should be the target of
future studies. For now we note that \citet{McCourt2015} found a reduced
mixing of accelerating clouds containing tangled magnetic fields.

We have argued that the magnetic field strength inside the stream changes its
magnetoacoustic response with the result that body modes have an increased
growth rate. For the cylindrical stream, we have seen that non-axisymmetric
surface modes have the dominant growth rate.  The value of the magnetic field
strength at the surface of the stream is therefore potentially more important
than its value inside the stream.  The associated magnetic field tension can
have the power to stabilize surface modes of the KHI.  While the magnetic
field might not have high values everywhere in the CGM, the magnetic field
strength could be highly enhanced at the stream surface due to magnetic
draping which inevitably occurs when an object is moving super-Alfv\'enically
through a magnetized medium (see e.g. \citealt{Dursi2008,
Pfrommer2010,McCourt2015}). Super-Alfv\'enic motion causes the ambient
magnetic field lines to accumulate at the stagnation point on the surface of a
cold cloud as the field is unable to adjust to the approaching body for
causality reasons. A steady state is reached once the magnetic energy density
of the draping layer matches the ram pressure of the incoming wind: the
dynamically important magnetic pressure pushes field lines over the body at
the rate of which new field lines are accumulated in the draping layer.
Provided the Alfv\'{e}n speed is larger than the gas velocity, the thin
magnetic draping layer that separates the two counter-flowing fluids can
suppress instabilities on scales significantly larger than its thickness
\citep{Dursi2007,Dursi2008}.

\section{Conclusion}
\label{sec:conclusion}

Star forming galaxies in the early Universe have been suggested to sustain
their star formation via filamentary accretion of cold gas from the cosmic web
\citep{Dekel2006,Dekel2009}.  A key question for this model is whether such
cold streams of gas can penetrate all the way down to the central galaxy,
where new fuel for star formation is needed, or whether the streams will be
disrupted and heated during their propagation through the hot, virialized halo
gas. Cosmological simulations have not yet attained the required spatial
resolution to address this question directly \citep{Sparre2019}, and such
simulations do not agree on whether the cold stream will survive the journey
through the CGM \citep{Nelson2013}.

This has motivated idealized studies of cold streams, such as the ones by
\citet{Mandelker2016}, \citet{Padnos2018} and \citet{Mandelker2019}, where
instability mechanisms can be more easily understood and the streams can be
properly resolved. These studies considered the cold streams using ideal
hydrodynamics. Here we extend the analysis to ideal MHD by introducing
magnetic fields in idealized simulations of cold streams feeding
massive halos in the early Universe.

We find using linear theory that a strong magnetic field ($\beta=1$) inside
the  stream changes its magnetoacoustic response and enhances the growth of
the KHI in the form of reflective waves that are primarily localized inside
the stream (so-called  reflective or body modes). While a strong magnetic
field has long been known to suppress surface modes of the KHI
\citep{Chandrasekhar1961}, a strong magnetic field can also enhance the
effective growth rate because body modes are an important instability
mechanism for a super sonic stream.

Although a magnetic field can enhance the growth rate of the KHI, we find that
a strong magnetic field outside the stream can make it difficult for the
stream to be disrupted in the non-linear phase of the KHI. This effect is
clearly seen with a uniform magnetic field, everywhere aligned with the
stream. In 2D, any perpendicular motion of the gas is associated with a
buildup of magnetic tension due to bending of magnetic field lines.  We show
in Fig.~\ref{fig:2D_snaps} how a stream is able to almost completely retain
its integrity when there is an ambient, uniform magnetic field with $\beta=1$.
 When the magnetic field strength is lower, $\beta=10$, the effect is lowered
enough that the stream breaks up. The magnetic field is however still strong
enough to significantly enhance the time scale on which the dense gas is fully
mixed with the CGM.

In 3D, the picture is slightly changed. The introduction of a third dimension
allows disturbances that vary around the surface of a cylindrical stream (and
not just along it, as for the 2D version). Since the magnetic field is assumed
to be aligned with the stream, these motions are not associated with any
magnetic tension. This means that azimuthal surface modes (disturbances
localized on the surface of the cylinder, so-called fluting modes) become
important in the more realistic 3D treatment. Additionally, and unlike
longitudinal surface modes, they are not stabilized by the supersonic flow.
The conclusion that azimuthal modes become dominant in 3D thus carries over
from the hydrodynamic case studied in \citet{Mandelker2019}.

As for the 2D case, the magnetic field is able to increase the growth rates of
the instabilities, again, it seems, by changing the magnetoacoustic response
of the stream. We find that the most unstable mode for our parameters is the
$m=4$ fluting mode  and that  the growth rate is increased by a factor $\sim
1.4$ with respect to the  hydrodynamic case for a strong magnetic field
($\beta=1$).

Although the effect is not as dramatic as in 2D, the magnetic field is also
able to suppress mixing of the cold stream with the ambient CGM in 3D (see
Fig.~\ref{fig:3D_snaps} for a visualization of the non-linear phases of the
KHI and Fig.~\ref{fig:dense-fraction-detailed} for a detailed comparison
between mixing in 2D and 3D). We find that magnetic fields with
$\beta^{-1}=0.1-1$ can increase the survival time of cold streams to be
$\sim$2--8 times longer than in our hydrodynamic simulations (see
Fig.~\ref{fig:disruption_sum} for a summary of the $\beta$-dependence in 3D).
The survival time can be used to estimate the minimum width a stream can have
without being disrupted before reaching the central galaxy. We find that
streams $\sim2-8$ times thinner can reach the central galaxy if the magnetic
field strength has $\beta^{-1}=0.1-1$, which corresponds to $\sim 0.3-0.8
\mu$G for characteristic CGM values of density and temperature (see
Section~\ref{sec:astro-impli} for details).

Our study has assumed a very idealized picture of a cold stream propagating
through the CGM. A more realistic treatment would include turbulent magnetic
fields in the CGM, radiative cooling, thermal conduction, self-gravity of the
stream \citep{Aung2019} and external gravity from the dark matter halo, which
leads to a varying CGM density as the stream penetrates deeper into the halo.
Nevertheless, our idealized model of a magnetized cold stream shows that
magnetic fields could potentially be very important for the dynamical
evolution of cold streams feeding galaxies at high redshift. This motivates
including magnetic fields in both future idealized simulations and in
cosmological simulations with enhanced CGM resolution.

\section*{Acknowledgments}
We thank the referee, Yuval Birnboim, for a detailed and insightful report
which helped us improve the manuscript. We thank Martin Sparre for sharing his
knowledge on CGM physics. TB and CP  acknowledge support by the European
Research Council under ERC-CoG grant CRAGSMAN-646955.

\bibliographystyle{mnras}
\bibliography{references}

\appendix

\section{Linear theory and code verification}
\label{sec:linear_theory}
The equations of ideal MHD \citep{Freidberg2014}, i.e.,
the mass continuity, momentum, induction
and entropy equations, are given by
\begin{align}
    \pder{\rho}{t} &= -\bs{\nabla} \bs{\cdot} \left(\rho \vec{\varv}\right) \ ,
    \label{eq:rho}
\end{align}
\begin{align}
    \rho \der{\vec{\varv}}{t} &=
    - \bs{\nabla} p
    - \bs{\nabla} \bs{\cdot} \left(\f{B^2}{2\mu_0}\mathbf{1} -
    \f{B^2}{\mu_0}\b\b \right) \ ,
    \label{eq:mom}
\end{align}
\begin{align}
    \pder{\vec{B}}{t} &= \bs{\nabla} \bs{\times} \left(\vec{\varv} \bs{\times} \vec{B}\right)
    \label{eq:ind}\ ,
\end{align}
\begin{align}
    \f{p}{\gamma -1}\der{\ln \left(p \rho^{-\gamma}\right)}{t} &=
    0 \ ,
    \label{eq:ent}
\end{align}
in SI units.
Here
$\bs{a}\bs{b}$ is the dyadic product of vectors $\bs{a}$ and
$\bs{b}$, $\rho$ is the mass density, $\vec{\varv}$ is the mean fluid
velocity, $p$ is the thermal pressure, $\vec{B}$ is the magnetic field
with local direction $\b$, $\mu_0$ is the magnetic permeability and
$\gamma=5/3$ is the adiabatic index.

The nonlinear dynamics of cold streams, including fragmentation and
mixing with the CGM, is studied in the main body of the paper by
solving Equations~\eqref{eq:rho}-\eqref{eq:ent} numerically with the aid
of the MHD code \textsc{Athena++} \citep{White2016}.

The linear dynamics of cold streams are governed by simpler, linearized
equations of ideal MHD. We present these linearized equations in 2D and 3D
in Sections~\ref{app:2D-lin} and
\ref{app:3D-lin} and solve them with the aid of \textsc{Psecas}
\citep{Berlok2019}.
The obtained linear solutions are then used in Section~\ref{sec:verification}
to verify that
our \textsc{Athena++} setup is able to accurately
capture the linear dynamics of the KHI.

\subsection{2D slab in Cartesian geometry}
\label{app:2D-lin}
\begin{figure}
\includegraphics[trim = 0 15 0 0]{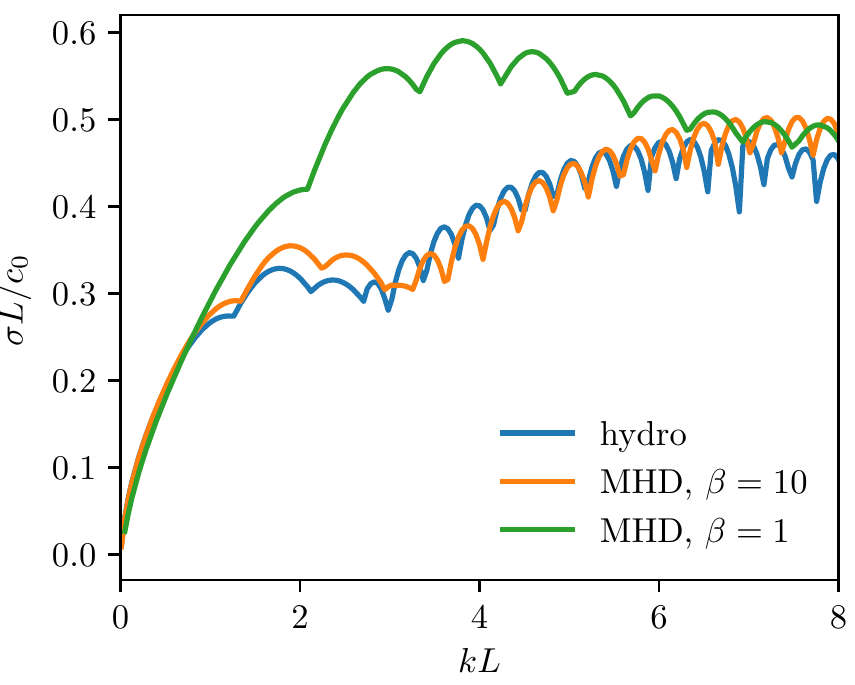}
\caption{Growth rates of the KHI in a dense and cold supersonic 2D slab which
  travels through an ambient dilute, hot medium. The KHI manifests itself as
  surface modes at low wave numbers and faster growing body modes at higher
  wavenumbers. The magnetic field, with strength characterized by the
  plasma-$\beta$, modifies the linear solutions and the phase velocity of waves.
}
\label{fig:theory-slab}
\end{figure}

For the magnetized KHI, the linearized equations have previously
been presented in Cartesian geometry \citep{Berlok2019}. For our specific
setup, where we
neglect heat conduction and viscosity as well as spatial variation in pressure
and magnetic field strength, Equations 17 to
21 in \citet{Berlok2019} reduce to
\begin{align}
    -\mathrm{i} \left(\omega - k \varv \right) \drho &=   - \mathrm{i} k  \dvx
    - \left(\der{\ln \rho}{z} + \pder{}{z} \right) \dvz \ ,
    \label{eq:rho-lin}
\end{align}
\begin{align}
    -\mathrm{i} \left(\omega - k \varv \right) \dvx &=
    - \pder{\varv}{z} \dvz
    - \mathrm{i} k \cs^2 \left(\drho + \dt\right) \ ,
    \label{eq:mom-x-lin}
\end{align}
\begin{align}
    -\mathrm{i} \left(\omega - k \varv \right) \dvz &=
    - \cs^2 \pder{}{z} \left(\drho + \dt\right)
    +\va^2\left[
    \pdder{}{z}
    - k^2
    \right] \dA \ ,
    \label{eq:mom-z-lin}
\end{align}
\begin{align}
    -\mathrm{i} \left(\omega - k \varv \right) \dA &= \dvz \ ,
    \label{eq:ind-lin}
\end{align}
\begin{align}
    -\mathrm{i} \left(\omega - k \varv \right) \dt
    &=  - \mathrm{i}k \f{2}{3} \dvx
    - \left(\der{\ln T}{z} + \f{2}{3} \pder{}{z} \right) \dvz \ .
    \label{eq:ent-lin}
\end{align}
Here $\delta A$ is the perturbed vector potential which is related to the
perturbed magnetic field by $\delta \vec{B} = \bs{\nabla}\bs{\times}(\delta A
\ey)$. The background profiles for $\varv$ and $\rho$ are given by
Equation~\eqref{eq:v-equi} and \eqref{eq:rho-equi}, respectively.
Equations~\eqref{eq:rho-lin} to
\eqref{eq:ent-lin}
is an eigenvalue problem, which we can solve with the rational Chebyshev
polynomial grid in \textsc{Psecas}. We assume as boundary conditions that
all perturbations approach zero as $z\rightarrow \pm\infty$.
We use this procedure to calculate the growth rate of the KHI in the
hydrodynamic limit ($\beta^{-1}=0$), for an intermediate field strength
($\beta=10$) and for a strong magnetic field ($\beta=1$).
The calculated growth rates are shown in Fig.~\ref{fig:theory-slab} as a
function of wavenumber. At low wavenumbers, where the KHI appears as surface
modes, the magnetic field suppresses the instability. This is in line with the
intuition obtained from incompressible theory \citep{Chandrasekhar1961}, who
found that the magnetic field appears as a surface tension term and suppresses
the instability. At higher wavenumber, the KHI appears as body modes. These
are waves that that have their dominant amplitude inside the stream
and grow in amplitude by tapping into the kinetic energy of the stream
\citep{Payne1985}.
For the body modes, the magnetic field increases
the growth rate of the KHI with respect to the hydrodynamic case.

\subsection{3D cylinder in polar coordinates}
\label{app:3D-lin}

\begin{figure*}
\includegraphics[trim = 0 20 0 0]{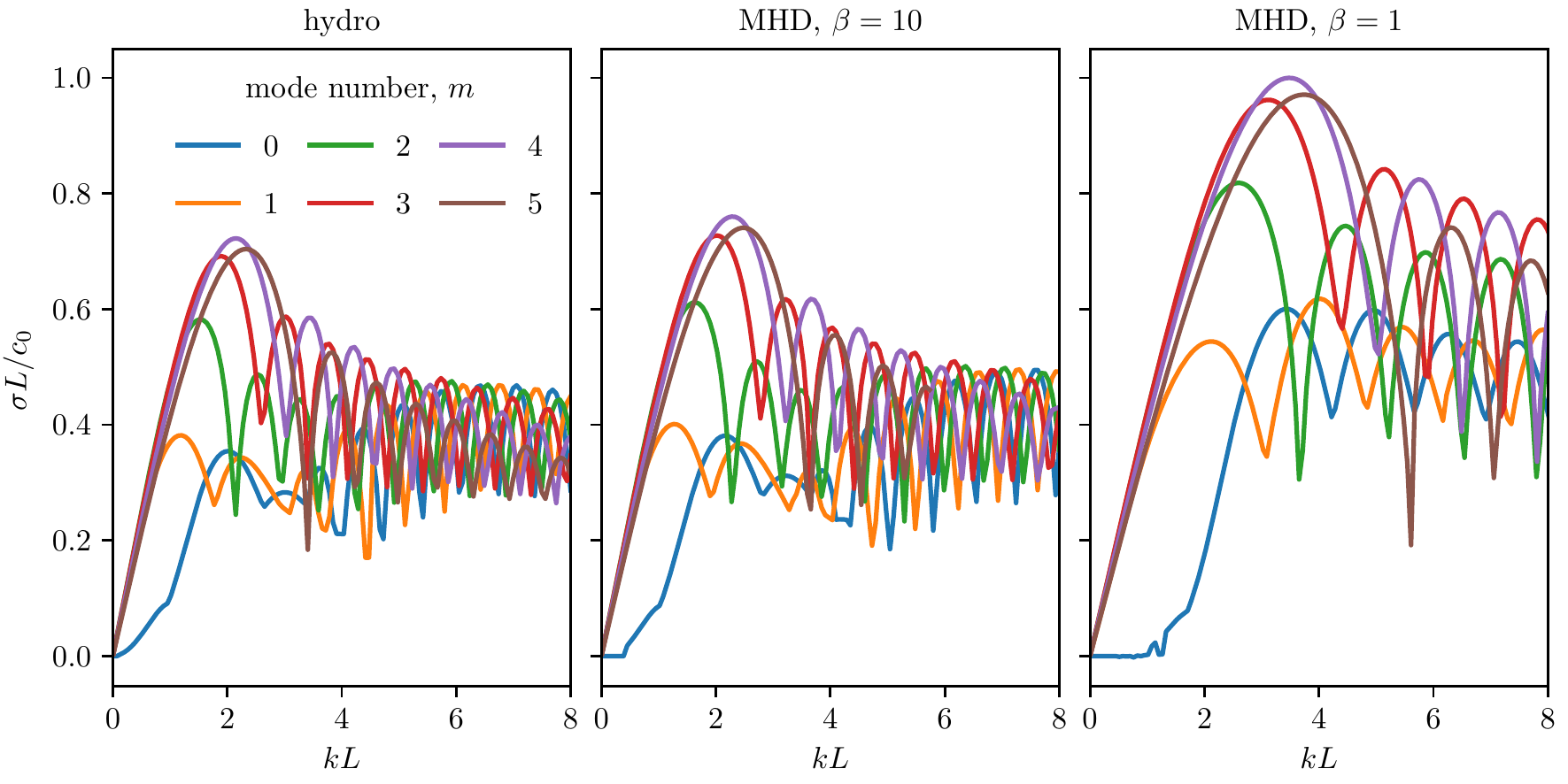}
\caption{Growth rates of the KHI in a dense, cold and supersonic 3D cylinder
  which travels through an ambient dilute, hot medium.}
\label{fig:theory-cylinder}
\end{figure*}
The linearized equations for the cylindrical stream with aligned
background magnetic field and velocity, i.e., $\vec{\varv}= \varv(r) \ez$
and $\vec{B}= B(r) \ez$, can be written in cylindrical coordinates as
\begin{align}
    -\mathrm{i}(\omega - k \varv) \delta \varv_r &=
    - \cs^2 \pder{}{r}\f{\delta p}{p}
    + \va^2 \left(\mathrm{i}k \delta b_r - \pder{\delta b_z}{r}\right)
    \label{eq:vr-lin}
    \ ,\\
    -\mathrm{i}(\omega - k \varv) \delta \varv_\phi &=
    - \cs^2 \f{\mathrm{i} m}{r}\f{\delta p}{p}
    -\va^2\left(\f{\mathrm{i} m}{r} \delta b_z - \mathrm{i}k \delta b_\phi \right)
    \ , \\
    -\mathrm{i}(\omega - k \varv) \delta v_z &=
    -  \cs^2 \mathrm{i} k \f{\delta p}{p}
    -   \pder{\varv}{r}\delta \varv_r  \ ,
\end{align}
\begin{align}
    -\mathrm{i}(\omega - k \varv) \f{\delta p}{p} &=
    - \gamma \mathcal{D}_r \delta \varv_r
    - \gamma \f{\mathrm{i} m}{r}\delta \varv_\phi
    - \gamma \mathrm{i} k \delta v_z \ ,
\end{align}
\begin{align}
    -\mathrm{i}(\omega - k \varv) \delta b_r &= \mathrm{i}k \delta \varv_r  \ , \\
    -\mathrm{i}(\omega - k \varv) \delta b_\phi &= \mathrm{i}k \delta \varv_\phi \, \\
    -\mathrm{i}(\omega - k \varv) \delta b_z &=
    - \mathcal{D}_r  \delta \varv_r + \pder{\varv}{r} \delta b_r
    -\f{\mathrm{i} m}{r} \delta \varv_\phi
    \label{eq:bz-cyl}
    \ ,
\end{align}
where the differential operator $\mathcal{D}_r$ is
\be
    \mathcal{D}_r \equiv \pder{}{r} + \f{1}{r} \ ,
\en
and $k$ and $m$ are the longitudinal and azimuthal wavenumbers, respectively.
The perturbation to the magnetic field $\delta \vec{B}$, is related to $\delta
\vec{b}$ by $\delta \vec{b} = \delta \vec{B}/B$ and we have used
the $\nabla \cdot \delta \vec{b}=0$ constraint,
\be
    \bm{\nabla} \bm{\cdot} \delta \vec{b} = \mathcal{D}_r \delta b_r +
        \f{\mathrm{i} m}{r} \delta b_\phi + \mathrm{i} k \delta b_z
        = 0 \ ,
\en
to simplify Equation~\eqref{eq:bz-cyl}. The background profiles for
 $\varv$ and $\rho$ are given in Equations~\eqref{eq:v-equi-cyl} and
\eqref{eq:rho-equi-cyl}, respectively.

We use \textsc{psecas} with the semi-infinite rational Chebyshev grid to solve
Equations \eqref{eq:vr-lin} to \eqref{eq:bz-cyl} on the semi-infinite interval
$r \in [0, \infty]$. This grid has the advantage that it is not necessary to
explicitly impose boundary conditions at $r=0$ and $r\rightarrow\infty$
\citep{Boyd1987a,Boyd}.  We present the growth rate as a function of
longitudinal wavenumber, $k$, for azimuthal wavenumbers $m=0$--$5$ in
Fig.~\ref{fig:theory-cylinder}. We again consider three different
magnetizations (hydro, $\beta=10$ and $\beta=1$) and find that the $\beta=1$
cylinder has the highest growth rates.  As in 2D this also appears to be
related to the presence of body modes. While the lower-order azimuthal modes
($m=0$ and 1) stretches across the radial extent of the stream, the faster
growing higher-order azimuthal modes exhibit a finite depth as measured from
the surface. With increasing azimuthal wave number $m$ the solutions
progressively attain the  character of surface modes (see
Fig.~\ref{fig:cylinder_modes}). We also find that the fastest growing
eigenmode has $m=4$, independent of magnetization. A compilation of the
fastest growing longitudinal wavenumber for each value of $m$ and $\beta=1$ is
presented in Table~\ref{tab:khi_table}. These are the modes shown in
Fig.~\ref{fig:cylinder_modes}.

\begin{table}
    \centering
    \caption{Fastest growing eigenmodes for the KHI in a cylindrical cold stream
      with diameter $L$, density $\rhos/\rho_0=50$, Mach number
      $\mathcal{M}_0=2$, smoothing length $a/L=0.05$ and $\beta=1$.}
    \label{tab:khi_table}
    \begin{tabular}{llll}
        \hline
        $m$ & $k_\mathrm{max} L$ & $\sigma L/c_0$ \\
        0 & 3.43372 & 0.60051 \\
        1 & 4.01473 & 0.61800 \\
        2 & 2.59792 & 0.81842 \\
        3 & 3.11557 & 0.96195 \\
        4 & 3.48196 & 0.99990 \\
        5 & 3.73777 & 0.97093 \\
        \hline
    \end{tabular}
\end{table}

\subsection{Code verification}
\label{sec:verification}

\begin{figure}
\includegraphics[trim = 0 25 0 0]{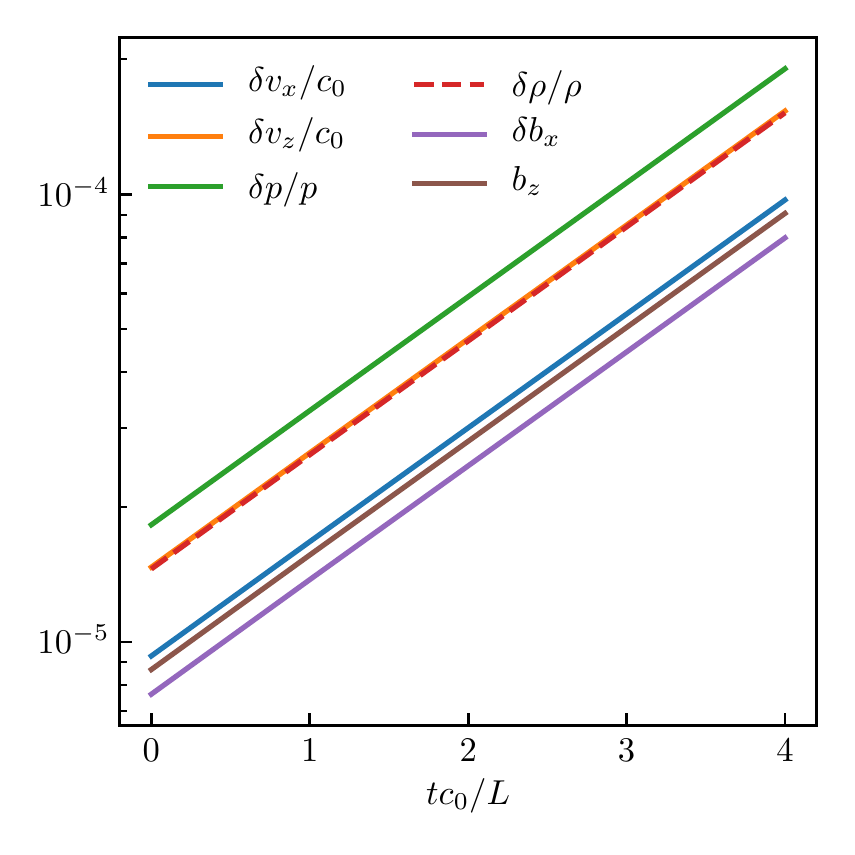}
\caption{Exponential growth of perturbations in a 2D
simulation with $\beta=1$.
The growth rate matches with the theory to within 1 \%.}
\label{fig:slab_beta1_fit}
\end{figure}

We use the linear solutions obtained in Sections~\ref{app:2D-lin} and
\ref{app:3D-lin} to verify our \textsc{Athena++} setup. We consider the
fastest growing mode in the $\beta=1$ simulations. For the 2D setup, this mode
has $k_\mathrm{max} L = 3.81069$, $\sigma L/c_0= 0.59086$ and is shown in
Fig.~\ref{fig:eigmode-slab}. Using the \textsc{psecas} solutions we seed all
physical variables in an \textsc{Athena++} simulation. Details of such a
procedure has been previously discussed in \citet{Berlok2019} for the double
periodic KHI of \citet{Lecoanet2016} where simulations where  initialized with
a Fourier series for each component of the linear solution. Here we instead
read in the  coefficients for the rational Chebyshev polynomial expansions of
the linear  solution and use those to construct the initial conditions in the
problem  generator in \textsc{Athena++}. This yields essentially perfect
exponential  growth of the instability, as evident in
Fig.~\ref{fig:slab_beta1_fit}. The  growth of perturbations is here tracked by
calculating the mean of the  absolute difference between the components and
their equilibrium values. The growth rate, measured by performing an
exponential fit, agrees to within  1 \% of the theoretical estimate. The
domain size of this simulation is $L_x\times L_z$ with  $L_x =
2\upi/k_\mathrm{max}$ and $L_z = 20 L_x$, i.e., it  is set up in a way such
that the fastest growing mode fits exactly in the  longitudinal direction. The
simulation has static refinement with 7 levels in  the perpendicular
direction, corresponding to a resolution of $\sim 310$ cells per stream width
inside the central $z\in [-L_x, L_x]$.

\begin{figure}
\includegraphics[trim = 0 25 0 0]{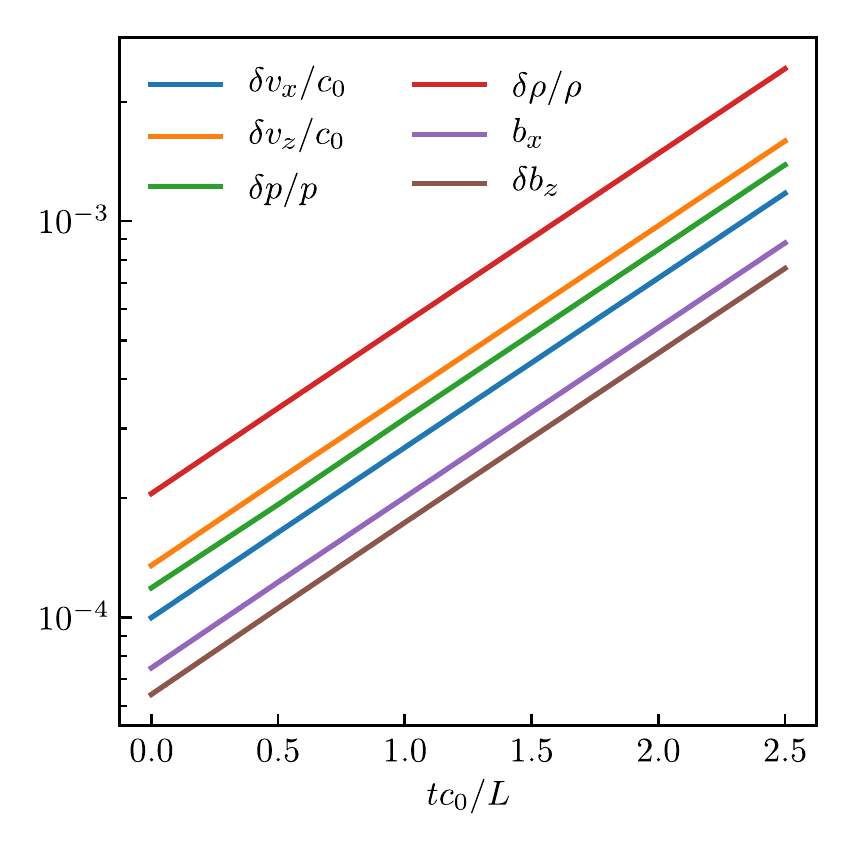}
\caption{Exponential growth of perturbations in a 3D simulation with
  $\beta=1$. The values for $\delta \varv_y$ and $\delta b_y$ are the same as
  values for $\delta \varv_x$ and $\delta b_x$ and are not shown to avoid clutter in
  the figure.  The growth rate matches with the theory to within 1.5 \%. }
\label{fig:cyl_beta1_fit}
\end{figure}

We repeat this procedure for a 3D simulation of the KHI in a cylindrical
cold stream. We seed the fastest growing $m=4$ mode with $\beta=1$ and
choose the domain size to exactly fit the mode and
to be much larger in the perpendicular direction, i.e.,
$L_z = 2\upi/k_\mathrm{max}$ and $L_x = L_y = 20 L_z$. We use the
coefficients of the linear solutions in terms of rational
Chebyshev polynomials on a semi-infinite domain to initialize the simulation.
The linear solutions are given in cylindrical coordinates while we use a
Cartesian grid in \textsc{Athena++}. We convert the solution from cylindrical
to Cartesian coordinates using a standard coordinate transformation,
i.e., $r = \sqrt{x^2 + y^2}$, $\phi = \mathrm{atan2}(y, x)$,
$\delta b_x = \cos (\phi) \,\delta b_r - \sin (\phi) \, \delta b_\phi$,
$\delta b_y = \sin (\phi) \,\delta b_r + \cos (\phi) \, \delta b_\phi$ and
similarly for $\delta \varv_x$ and $\delta \varv_y$.
The magnetic field is initialized directly
and not via a magnetic vector potential. This introduces a numerical divergence
of the magnetic field, which could potentially become a problem for the
constrained transport scheme in \textsc{Athena++}. The direct initialization
however suffices for this linear test, and we obtain excellent exponential
growth, see Fig.~\ref{fig:cyl_beta1_fit}. The 3D simulation has 6 levels of
static refinement and $\sim 140$ cells per stream width inside the central
$(x, y) \in [-L_z, L_z]\times[-L_z, L_z]$.

\begin{figure*}
\includegraphics[trim = 0 28 0 0]{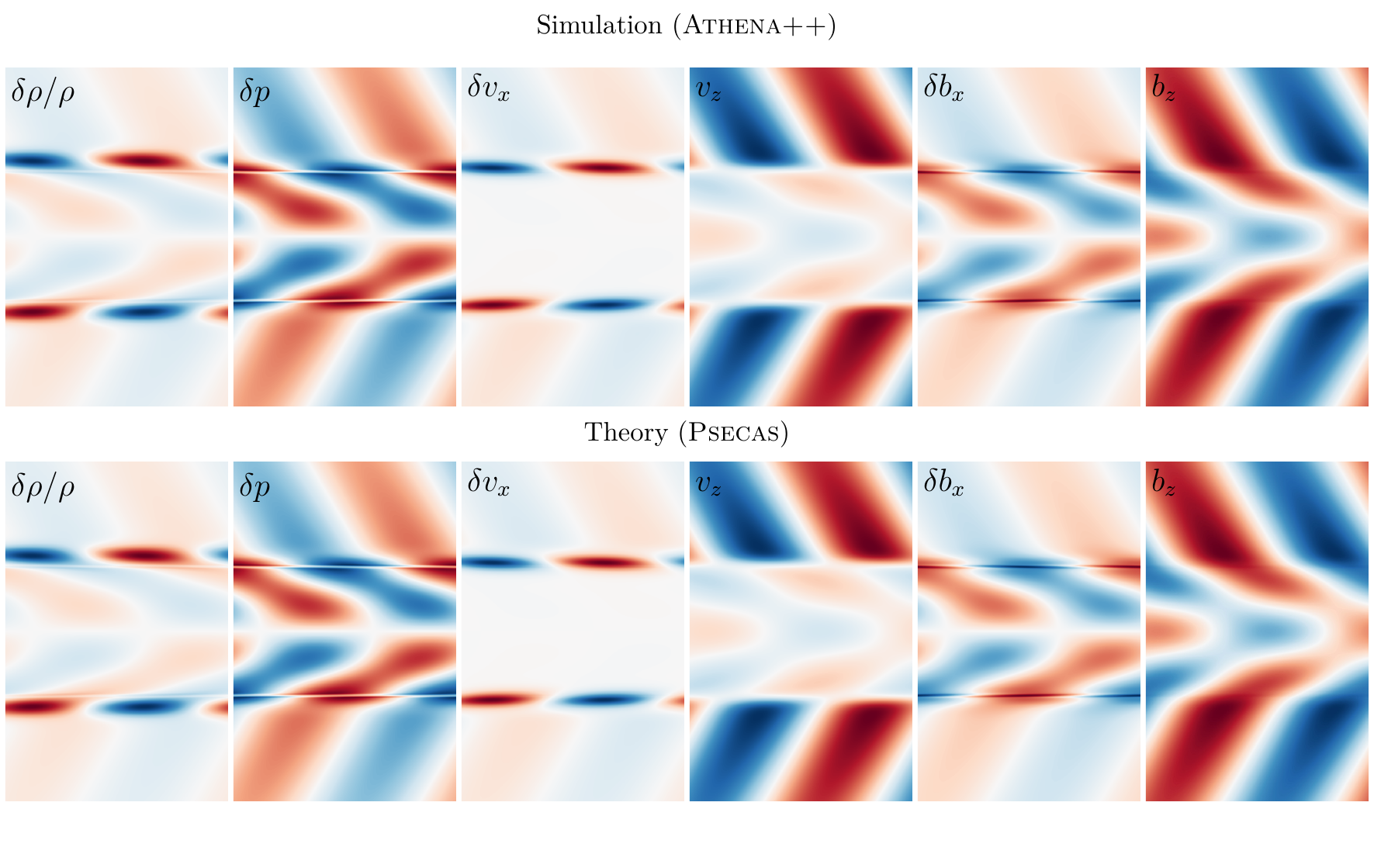}
\caption{Comparison between eigenmodes in a 2D \textsc{Athena++} simulation
(upper row) and the linear theory (lower row) at $t= 4 L/c_0$.}
\label{fig:slab-eigmodes}
\includegraphics[trim = 0 28 0 0]{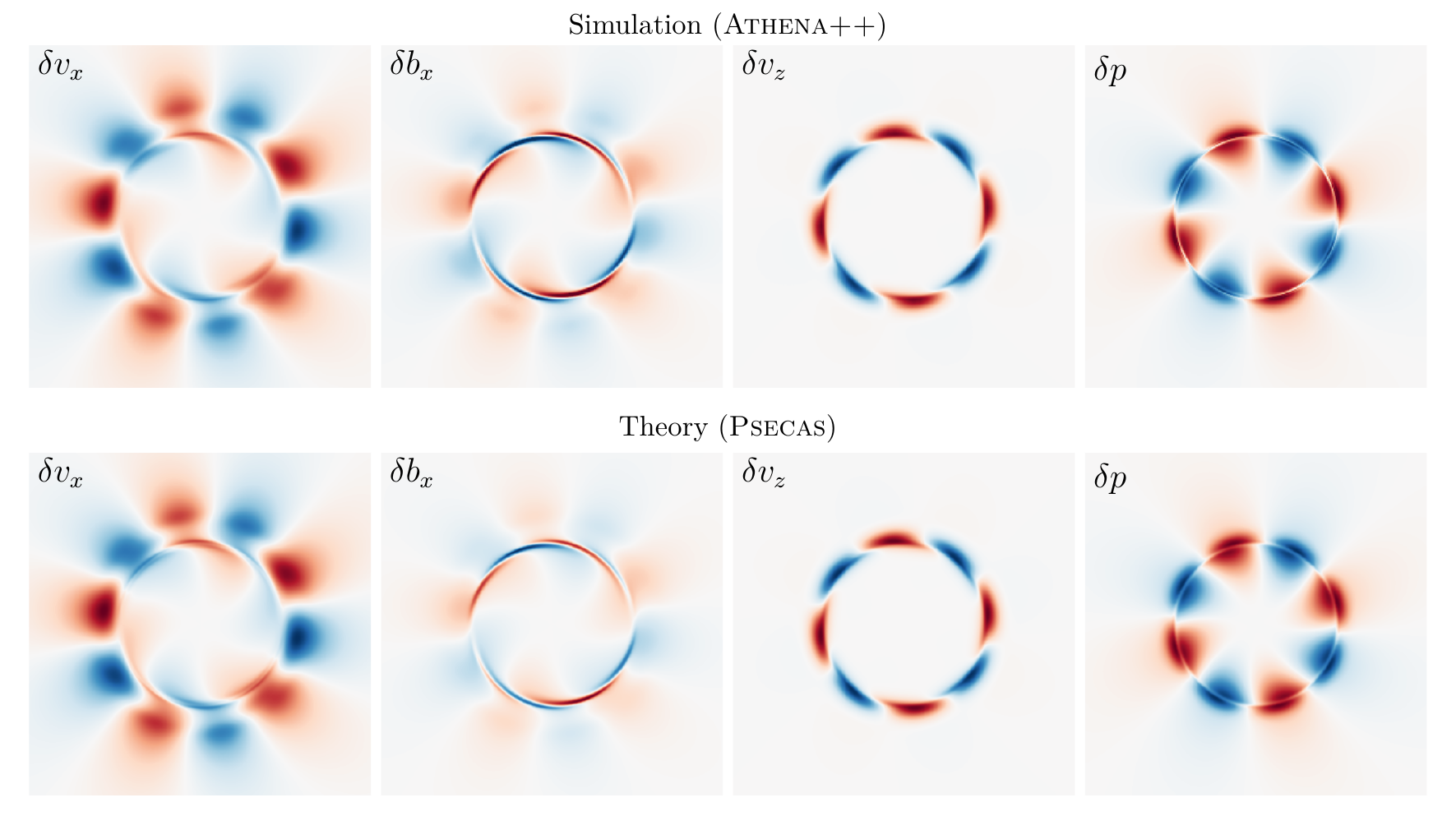}
\caption{Comparison between select eigenmodes in a 3D \textsc{Athena++} simulation
(upper row) and the linear theory (lower row) at $t= 2.5 L/c_0$.
We show $xy$-slices of the domain $(x, y) = [-L, L]\times[-L, L]$ at $z=0$.
The upper and lower row use the same color scales.}
\label{fig:cylinder-eigmodes}
\end{figure*}

We have also verified that the deviations from
the background in the simulations agree with the eigenmodes obtained
with \textsc{psecas}. We present the background deviations after they have
grown by a factor of $\sim 10$ in amplitude in the top rows of
Figs.~\ref{fig:slab-eigmodes} and \ref{fig:cylinder-eigmodes}. These are
almost indistinguishable from the \textsc{psecas} eigenmodes which are shown
in the lower rows of Figs.~\ref{fig:slab-eigmodes} and
\ref{fig:cylinder-eigmodes}. We have calculated the discrepancy, using as a
measure the mean of the absolute difference between simulation and theory
inside the domain shown in the figures.
This number is $\sim 1$ \% of the amplitude of the eigenmodes.
The tests verify that our \textsc{Athena++} setups are able to accurately
model the supersonic version of the magnetized KHI.

\subsection{Parameter study of 2D slab}
\label{app:parameter-study}

\begin{figure*}
\includegraphics[trim= 0 25 0 0]{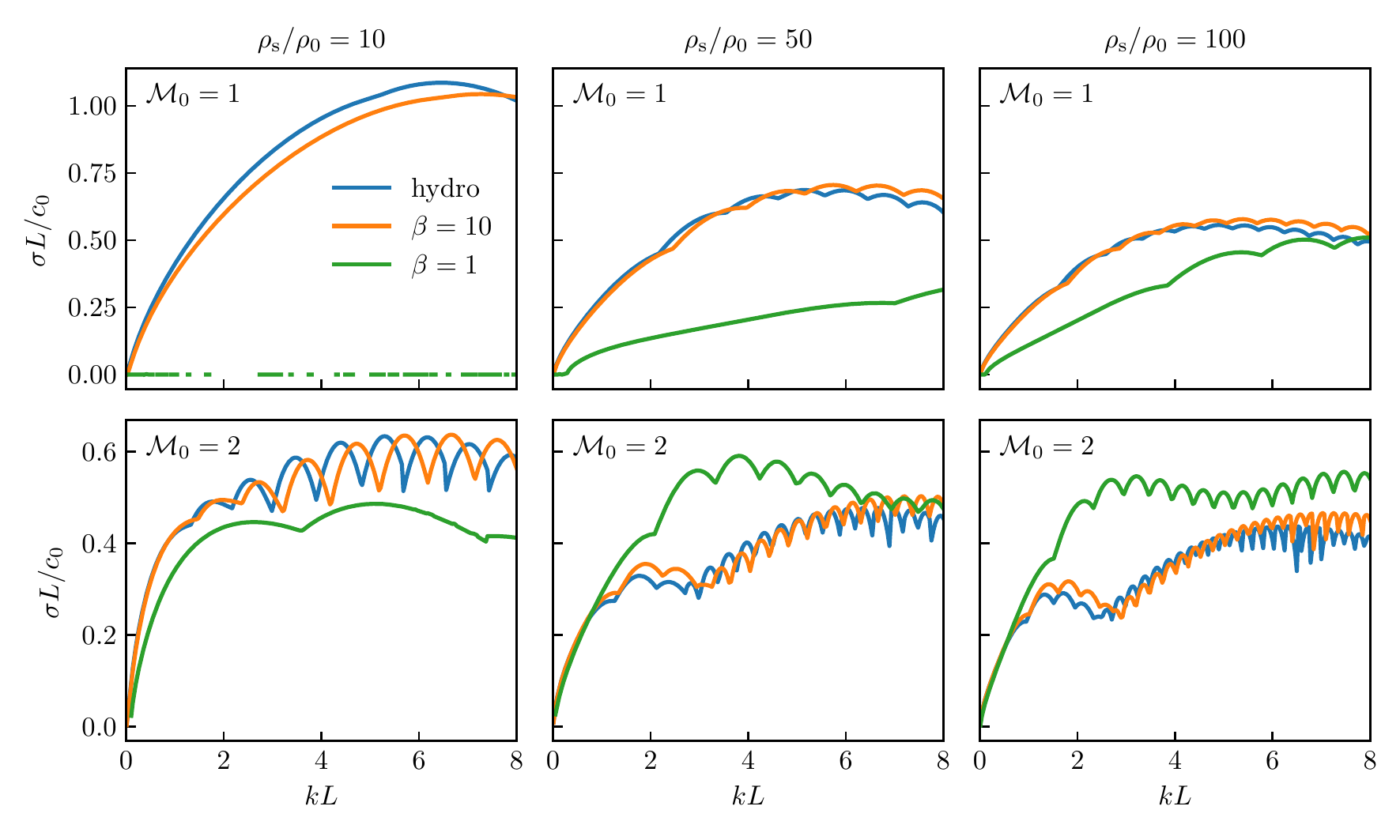}
\caption{Growth rates for the 2D slab stream as a function of wavenumber.
    The upper (lower) row of panels has $\mathcal{M}_0=1$ ($\mathcal{M}_0=2$)
    and the left, middle and right columns have $\delta=9$, $\delta=49$ and
    $\delta=99$. We compare hydrodynamic and magnetized calculations with
    $\beta=1$ and 10. Magnetic fields generally suppress the KHI at low $\delta$
    and $\mathcal{M}_0$ but enhance the growth rate at high $\delta$ and
    $\mathcal{M}_0$. Magnetic fields also lead to a change in the location of
    resonant peaks in the growth rate of body modes.}
\label{fig:parameter_study}
\end{figure*}

We present a parameter study for the 2D slab stream. The main
text considers a dense, cold stream with
$\mathcal{M}_0=2$ and $\rhos/\rho_0=50$ (that is,
$\delta=49$) and compares a hydrodynamic study with
MHD studies with $\beta=1$ and 10.
In Fig.~\ref{fig:parameter_study} we consider the same magnetizations
but extend the study to three different density contrasts ($\delta=9$, 49
and 99) and two different Mach numbers ($\mathcal{M}_0=1$ and 2).
In this figure, the middle panel in the lower row
is identical to Fig.~\ref{fig:theory-slab}.

We find that magnetic fields suppress the KHI at low density contrasts and
flow speeds. For $\mathcal{M}_0=1$, $\beta=1$ leads to full suppression when
$\rhos/\rho_0=10$ but only to inhibited growth when $\rhos/\rho_0=50$ or 100.
At higher flow speeds, $\mathcal{M}_0=2$, the $\beta=1$ growth rate is
decreased with respect to the hydrodynamic case for $\rhos/\rho_0=10$ only. At
higher density contrasts, $\rhos/\rho_0=50$ and 100, the $\beta=1$ calculation
instead yields growth rates that are \emph{increased} with respect to the
hydrodynamic case. The magnetic field-induced increase in growth rate is only
found at high wavenumbers where the KHI can occur as body modes. At low
wavenumbers, where the KHI appears as surface modes, the magnetic field always
inhibits the instability via magnetic tension. Hydrodynamic body
modes become dominant when $\mathcal{M}_0$ and $\delta$ are
large \citep{Mandelker2016}, and this is also the regime where MHD body
modes become important. Hydrodynamic body modes have a growth rate which
depends on the sound wave speeds inside and outside the stream
(see equation H8 in \citealt{Mandelker2016}).
We therefore believe the increase in growth rate of body modes occurs
because the magnetic field
increases the phase speed of waves compared with hydrodynamic sound waves.
The dispersion relations for a magnetized slab presented in
\citet{Hardee1992} and for a magnetized cylinder presented in
\citet{Appl1992}, although derived in the vortex sheet approximation
and with under-dense jets in mind,
might be able to provide additional insights into the behavior of supersonic
cold streams and help with this interpretation of our findings.
We leave such comparisons for future studies.

Finally, we find that the wavelength at which the body modes
are resonant changes with magnetic field strength. This is particularly
evident in the solutions for $\beta=10$ and hydrodynamics
with $\rhos/\rho_0=10$, $\mathcal{M}_0=2$ (the lower left panel) where the
$\beta=10$ resonant peaks are slightly shifted to the right. We again speculate
that these shifts are due to the difference in propagation speeds of waves in
unmagnetized and magnetized fluids.

\bsp	
\label{lastpage}
\end{document}